\documentclass[%
 reprint,
amsmath,amssymb, aps,
]{revtex4-2}

\usepackage[english]{babel}
\usepackage[utf8]{inputenc}
\usepackage[colorinlistoftodos, color=blue!40, prependcaption]{todonotes}
\usepackage{orcidlink}
\usepackage{url}

\usepackage{amsthm}
\usepackage{mathtools}
\usepackage{physics}
\usepackage{xcolor}
\usepackage{graphicx}
\usepackage[left=17mm,right=17mm,top=20mm,columnsep=15pt]{geometry} 
\usepackage{adjustbox}
\usepackage{placeins}
\usepackage[T1]{fontenc}
\usepackage{lipsum}
\usepackage{csquotes}
\usepackage[percent]{overpic}

\usepackage{multirow}
\usepackage{hyperref}
\hypersetup{
    colorlinks=true,
    linkcolor=blue,
    citecolor = blue,
    filecolor=magenta,      
    urlcolor=blue,
    pdftitle={Overleaf Example},
    pdfpagemode=FullScreen,
    }
\begin{document}

\title{Probing the Type 3 interacting dark-energy model using matter pairwise velocity}

\newcommand{\inst}[1]{\textsuperscript{#1}}

\author{
Kin Ho Luo\inst{1} \orcidlink{0009-0009-1698-700X}}
\email{ 1155129240@link.cuhk.edu.hk}
\author{Ming-chung Chu\inst{1} \orcidlink{0000-0002-1971-0403}}
\email{mcchu@phy.cuhk.edu.hk}
\author{Wangzheng Zhang\inst{1, 2} \orcidlink{0000-0003-0102-1543}}
\email{zhang@iap.fr}

\affiliation{
\inst{1} Department of Physics, The Chinese University of Hong Kong, Sha Tin, N.T., Hong Kong\\
\inst{2} Institut d'Astrophysique de Paris, UMR~7095, CNRS, Sorbonne Universit\'e, 98~bis~boulevard Arago, 75014~Paris, France
}

\date{\today} 

\begin{abstract}
Dark sector interactions can be explored via the so-called Type 3 model where dark matter and dark energy exchange momentum only, so as to minimize deviations from the $\Lambda$CDM background expansion history. Using N-body simulations, we analyze the imprint of Type 3 model parameters, the momentum exchange coupling constant $\beta$ and the slope of scalar field potential $\lambda$, on large-scale structure observables, particularly the matter pairwise velocity statistics. We find that the effects of $\beta$ ($<0$) and $\lambda$ on the mean matter peculiar pairwise velocity and velocity dispersion are degenerate. Our results highlight the potential of velocity statistics as a probe of dark sector interactions and underscore the importance of disentangling $\beta$ and $\lambda$ in cosmological analyses.


\end{abstract}


\maketitle

\section{Introduction} \label{sec:introduction}
Observations of Type Ia supernovae \cite{Perlmutter1999,riess2009}, baryon acoustic oscillations (BAO) \cite{Eisenstein2005}, cosmic microwave background (CMB) \cite{Aghanim2020}, and galaxy cluster mass functions \cite{Vikhlinin2009} provide robust evidence that the Universe is in an epoch of accelerated expansion, driven by dark energy (DE). There are various models to account for DE, such as modified gravity \cite{COPELAND2006} and a dynamical scalar field called quintessence \cite{Clifton2012}, but it is most commonly represented as a cosmological constant $\Lambda$, which, in conjunction with cold dark matter (CDM), are the dominant energy components of the standard cosmological model ($\Lambda$CDM) in the late-time Universe. While $\Lambda$CDM remains statistically favored for fitting most datasets \cite{Bennett2013, Aghanim2020, lampeitl2010, riess2009}, it faces significant tensions: the Hubble tension between values of the Hubble constant derived from CMB ($H_{0}$) and local measurements, such as Type Ia supernovae \cite{dhawan2018}; and the $S_8$ tension, where CMB-predicted matter clustering exceeds values from cosmic shear surveys by 2–3$\sigma$ \cite{DES2022}.

Furthermore, the $\Lambda$CDM model does not provide adequate explanations for the \textit{coincidence} and \textit{fine-tuning} problems \cite{baumann2022}. Recent BAO measurements from the Dark Energy Spectroscopic Instrument (DESI) further challenge the model, suggesting new physics in the dark sector. Specifically, the data hint at a dynamical equation of state (EOS) for DE, which could be indicative of nongravitational interactions within the dark sector or other extensions of the standard model \cite{Adame2025a, Adame2025b, Adame2025c}. To address these issues, a plethora of alternative models have been proposed \cite{dolgov2003, troester2021, mancini2019, joudaki2017, di_valentino2020b, di_valentino2021}. A promising example is the interacting dark-energy (IDE) models, which allow DE and dark matter (DM) to interact \cite{amendola2000, tamanini2015, pourtsidou2013}. These models seem quite natural, as they can explain cosmic coincidence such that $\Omega_{\Lambda, 0}/\Omega_{DM, 0} \sim \mathcal{O}(1)$. The relevant mechanisms seem to offer alleviation of the $H_0$ and $S_8$ tensions \cite{pourtsidou2016, di_valentino2021b}. However, many of these models exhibit noticeable shortcomings, such as large quantum corrections that can exacerbate the cosmological constant problem \cite{damico2016, marsh2017} and \textit{ad hoc}, purely phenomenological coupling parameters that fit CMB data worse than $\Lambda$CDM's \cite{GomezValent2020, Bean2008, Xia2009}. In most IDE models, there is an energy exchange between the two components in the dark sector, which can modify the background evolution and lead to nonadiabatic instability in dark sector perturbations \cite{valiviita2008}.

Numerous N-body simulations have been conducted on IDE models associated with energy transfer \cite{maccio2004, baldi2010,baldi2012, zhang2018, liu2022}, and most resulted in modifications to the matter power spectrum and halo mass function. By treating DE as a scalar field and DM as a fluid, it was found that the inner DM halo density decreases with respect to $\Lambda$CDM's, with increased suppression for stronger couplings \cite{baldi2009}. Similar effects on the inner (but not the outer) DM halo density profile were also found in \cite{li2011}, but with reduced suppression for stronger couplings relative to $\Lambda$CDM's, the exact reason for which remains unexplored. Additionally, the nonlinear matter power spectrum exhibits enhancement across all scales, with this effect becoming prominent at early redshift ($z$ = 49). This enhancement arises because the coupling between DE and DM modifies the growth of density perturbations, leaving an imprint on structure formation. As a result, the initial conditions for N-body simulations need to be modified relative to those used in the $\Lambda$CDM framework. A recent study by \cite{zhao2023} investigated the halo concentration-mass (\textit{c-M}) relation for a class of IDE models. The correlation between the \textit{c-M} relation and the energy interaction coupling strength $\xi_2$ is shown to be well fitted by a power law and so the former serves as a sensitive proxy for $\xi_2$. 

In the last decade, N-body simulations have been extended to IDE models associated with momentum transfer only \cite{pourtsidou2013, pourtsidou2016, skordis2015, simpson2010, kase2020, kase2020b, linton2022,chamings2020}. In \cite{baldi2016}, simulations were conducted based on the dark scattering model. Given DE and the nonrelativistic velocities of DM, the dark sector is modeled as two fluids undergoing elastic scattering similar to Thomson scattering. The DM then experience an additional drag term from the scattering, so the structure formation is affected. By employing a dynamical DE EOS $\omega$ that approaches the cosmological constant EOS $\omega_\Lambda = -1$ at late times, it was found that, at $z=0$, the matter power spectrum is suppressed by approximately 10\% at linear scales. This suppression diminishes on nonlinear scales. As a result, the suppressed power spectrum contributes less to weak lensing signals, leading to a lower $\sigma_8$ value from low-redshift probes, potentially alleviating the $S_8$ tension. This work was extended in \cite{bose2018} to test models with different $\omega$ and momentum interactions. It was concluded that linear perturbations increase with the elastic scattering cross section $\xi$. Additionally, it was observed that the power spectrum amplitude is enhanced (suppressed) at larger (smaller) scales for the constant quintessence case $\omega = -0.9$ relative to the $\Lambda$CDM's, and the effects are reversed for the phantom case $\omega = -1.1$. 

Yet another interesting class of the scattering model demonstrates its potential to probe nongravitational interactions of DE by leaving the imprints of scattering between DE and baryons on nonlinear structure formation, with subtle effects on the CMB. In \cite{ferlito2022}, the momentum exchange solely between DE and baryons is considered. Depending on whether the DE component is in a quintessence or phantom-like regime, its interaction imparts an additional cosmological friction (repulsive) or drag (attractive) force to baryonic particles, respectively. The work does not explicitly model nonadiabatic processes, such as radiative cooling and active galactic nuclei (AGNs) feedback, to concentrate on the signatures of DE baryons scattering in the observables of interest. They observed noticeable enhancement (suppression) of the nonlinear power spectrum and the inner baryon density profile relative to $\Lambda$CDM's for the quintessence- (phantom-) like cases at the nonlinear scales, the exact opposite of the measurements at the linear scales, where the baryons' peculiar velocities line up with the gravitational potential gradients, meaning that baryon scattering acts in the same direction as the gravitational force. However, this breaks down at the nonlinear scales due to the growing importance of angular momentum in collapsed structures. In this regime, for the quintessence (phantom) case, DE-baryons scattering acts as a cosmological friction (drag) force that facilitates (opposes) the loss of angular momentum by baryonic structures, altering the virial equilibrium of bound structures, causing them to contract (expand), which ultimately increases (decreases) the efficiency of nonlinear structure formation.

Recently, a new class of pure momentum transfer models has attracted significant interest: the Type 3 model \cite{pourtsidou2013}. This model addresses the shortcomings of IDE models by considering momentum transfer within the dark sector only \cite{pourtsidou2013}, parameterizing momentum exchange with a coupling constant $\beta$ and the DE EOS with the slope of scalar field potential $\lambda$. For certain combinations of $\beta$ and $\lambda$, this model leaves the background expansion history unchanged with respect to that of the $\Lambda$CDM model, achieving an excellent fit with the CMB data \cite{pourtsidou2016}. With a negative $\beta$ within the dark sector, it has even been shown to alleviate the $S_8$ tension \cite{pourtsidou2016, chamings2020}. Furthermore, the Type 3 model introduces DE-DM couplings at the level of the action, providing a more intuitive and direct insight into the interactions and their impact on the background expansion and structure formation. Because this model is described by a Lagrangian formalism, instabilities can be easily determined and avoided \cite{chamings2020}.

While the CMB effectively constrains $\lambda$ through the background cosmology, the growth of large-scale structure (LSS) offers a unique sensitivity to $\beta$, as the momentum exchange effect directly leaves imprints on the structure formation. Unlike the CMB, which contains significant early-Universe information alongside late-time contributions, LSS provides more direct probes of the late-time gravitational collapse and velocity field, where deviations from $\Lambda$CDM due to dark sector coupling become pronounced.

The effects of the Type 3 model on background perturbations and structure formation can be qualitatively described using linear perturbation theory tools such as \texttt{CLASS} \cite{lesgourgues2011} and \texttt{CAMB} \cite{lewis2000, howlett2012}. However, a comprehensive understanding of these effects requires an investigation into the nonlinear evolution of LSS through N-body simulations.

In \cite{palma2023}, N-body simulations were conducted to study the nonlinear effects of the Type 3 model from \cite{pourtsidou2013}. For positive values of $\beta$, the matter power spectrum is suppressed at nonlinear scales, contrary to the monotonic enhancement predicted by linear numerical simulations by \texttt{CLASS} at all scales. This occurs because the DM particles in bound structures experience reduced cosmological friction and retain more angular momentum than in the uncoupled scenario. This alters the virial equilibrium of the collapsed structures, resulting in their slower contraction and collapse compared to the uncoupled case. Consequently, the nonlinear structure formation and power spectrum are suppressed.  
Considering the new physics that the Type 3 model may reveal, it is fruitful to explore the relationships between its parameters, $\beta$ and $\lambda$, and other less-studied halo properties. 

Notably, the matter velocity field
 contains half of the phase-space information, yet it has not been used in previous work to constrain IDE models. In this paper, we study the mean peculiar pairwise velocity $v_{12}(r)$, or pairwise velocity in short,  and its dispersion $\sigma_{12}(r)$, which are defined as the mean and standard deviation of the relative peculiar velocities between pairs of objects projected along their lines of separation, respectively. The mean peculiar pairwise velocity can be used to infer the kinematic Sunyaev-Zeldovich effects \cite{bhattacharya2008,calafut2021} and constrain the local growth rate $f\sigma_8$ \cite{howlett2017, dupuy2019} as well as the cosmological parameters $\Omega_{m}$ and $\sigma_8$ \cite{juszkiewicz2000, feldman2003, ma2015, 2025ApJ...978L...6Z}, and more recently to probe the effects of sterile neutrinos on the LSS \cite{Hu_2025}. The mean peculiar pairwise velocity dispersion $\sigma_{12}(r)$ has been used to probe the properties of star formation \cite{tinker2007, vandenbosch2007, loveday2017}, constrain $\Omega_{m}$ \cite{davis1983, jing2002, zehavi2002, hawkins2003}, and measure the neutrino mass and chemical potential \cite{zhang2024}. Together, they serve as powerful tools for examining and constraining cosmological models.

In this paper, we study the dependences of $v_{12}(r)$ and  $\sigma_{12}(r)$ on the two parameters of the Type 3 model, $\beta$ and $\lambda$. 
The organization of this paper is as follows. Section \ref{sec: theory and sim} details the theoretical formalism of the Type 3 model and its implementation in N-body simulations. Section \ref{sec:velocity} provides the background of the mean peculiar pairwise velocity $v_{12}(r)$ and the velocity dispersion $\sigma_{12}(r)$. Section \ref{sec:results} presents the results, and Section \ref{sec:conclusions} summarizes and concludes the paper. 

\section{Type 3 model and simulations} 
\label{sec: theory and sim}
\subsection{Lagrangian and field equations}
\label{sec:develop}

It was shown by \cite{bahamonde2018} that varying the value of $\lambda$ leads to different stable solutions for the evolution of scalar field density. Specifically, for $\lambda^2 < 2 $, the solution resembles that of an accelerating Universe, driven by a sufficiently flat scalar field potential, corresponding to an effective equation of state $\omega_{eff} = \lambda^2/3 - 1$. Additionally, dynamical analyses of IDE models involving energy and momentum transfer have shown similar stable solutions across different ranges of $\lambda$ \cite{amendola2020, liu2024}. In summary, $\beta$ and $\lambda$ modify the kinetic and potential energy of the scalar field, respectively.

The defining characteristic of the Type 3 model of \cite{pourtsidou2013} is that the energy coupling between DE and DM, represented by the coupling current $ J_{\mu} $, vanishes at the background level in the energy conservation equation (see Appendix \ref{sec: equations}). This ensures no energy exchange between the dark sector components for all combinations of $ \beta < 1/2 $ and $ \lambda $.



Following \cite{pourtsidou2013}, the Lagrangian of the Type 3 model can be written as
\begin{equation} L = F(Y,Z,\phi) + f(n_c) , \label{ori_lag} \end{equation}
where $\phi$ is the DE scalar field, $Y=\frac{1}{2}(\nabla_\mu{\phi})^{2}$ is the kinetic energy density of the scalar field, $n_c$ is the DM number density, and $Z= u^{\mu}\nabla_\mu{\phi}$ represents the coupling of the four-velocity of the DM fluid $u^{\mu}$ to the gradient of the scalar field that gives rise to momentum transfer in the dark sector. Here, $F$ is the scalar field Lagrangian governing dark energy dynamics, while $f$ is the uncoupled DM fluid Lagrangian.

Based on Eq.~\eqref{ori_lag}, we assume a quintessence scalar field, with a fixed speed of sound $c_{s}=1$, and that the coupling takes a quadratic form, giving the scalar field Lagrangian 
\begin{equation} F = Y + \beta Z^{2} + V(\phi). \label{scf_lag} \end{equation}

Here, $\beta$ is the momentum coupling constant, and the scalar field potential is $V(\phi)=V_{0}e^{-\lambda\kappa\phi}$, where $\kappa = m_P^{-1}=\sqrt{8\pi G}$ is the inverse reduced Planck mass, the scalar field $\phi$ is in units of $m_P$ and this form is well-motivated by early-time inflation.

Focusing on the DM frame and defining the modified metric tensor $ \tilde{g}^{\mu\nu} = g^{\mu\nu} + 2\beta u^{\mu}u^{\nu} $, it follows from Eq.~\eqref{scf_lag} that the action of the scalar field takes the form
\begin{eqnarray} \nonumber  S_\phi = - \int d^4x \sqrt{-g} \left[ \frac{1}{2} \tilde{g}^{\mu\nu} \phi_{\mu} \phi_{\nu} + V(\phi) \right] \\ \to  \int dt d^3x a^3 \left[ \frac{1}{2} (1 - 2\beta) \dot{\phi}^2 - \frac{1}{2} |\vec{\nabla} \phi|^2 - V(\phi) \right].\label{eq:lagrangian} \end{eqnarray}

It should be noted that the first term of Eq.~\eqref{eq:lagrangian} inside the bracket becomes negative for $\beta>0.5$. It corresponds to a "ghost" scalar field with field amplitude and perturbations growing without bound and is therefore unrealistic. 
 
The variation of the action allows the energy-momentum tensor of the scalar field $T^{(\phi)}_{\mu\nu}$ to be separated from the DM fluid's, giving 

\begin{equation}
T^{(\phi)}_{\mu \nu} = F_Y \phi_\mu \phi_\nu - F g_{\mu \nu} - Z F_Zu_\mu u_\nu, \label{Energy_Tensor}
\end{equation}

where $F_Y$ and $F_Z$ denote the derivatives of the quintessence function with respect to $Y$ and $Z$.

\subsection{Background evolutions and perturbations}
Given Eq.~\eqref{scf_lag}, we obtain the Klein-Gordon equation describing the scalar field continuity evolution \cite{pourtsidou2016}, 
\begin{equation}
\ddot{\phi} + 2{\cal{H}\dot{\phi}} + \left( \frac{1}{1 - 2\beta} \right) a^2 \frac{dV}{d\phi} = 0, 
\label{eq: scf expli equation}
\end{equation}
where the dot denotes the derivative with respect to conformal time and $\cal{H}$ is the conformal Hubble parameter.

Similarly, the continuity equation of the DM fluid is
\begin{equation}
    \dot{\rho}_{c} + 3{\cal{H}}{\rho_{c}} = 0. \label{evolution}
\end{equation} 

Here, $\rho_{c}=\rho_{c,0}a^{-3}$ denotes the DM fluid energy density, and $\rho_{c,0}$ is the energy density at $a=1$.

To describe the effects of the Type 3 model on the LSS, we consider the linear perturbations of the flat Friedmann-Robertson-Walker metric in the Newtonian gauge for the ease of physical interpretation and implementation in our N-body simulations, which is given by \cite{baumann2022}
\begin{equation}ds^2 = a^2(\eta) \bigl( -\left( 1+ 2\Psi \right) d\eta^2 + \left( 1- 2\Phi  \right)  dx^i dx_i \bigl),  \end{equation}
\\
where  $i = 1, 2, 3$ stands for the spatial index, $\eta$ is the conformal time, and $\Psi$ and $\Phi$ are the scalar perturbations, with the former term playing the role of the gravitational potential in the Newtonian limit.

Applying Eq.~\eqref{evolution}, the evolution of the DM fluid perturbation is \cite{baumann2022} 
\begin{equation}
    \dot{\delta}_{c}= -\theta_{c} + 3\dot{\Phi} \label{perturbations},
\end{equation}

in which ${\delta}_{c} = \frac{{\delta{\rho}_{c}}}{{{\rho}}_{c}}$ is the density contrast, $\theta_{c} = \nabla_{i} {{v}_{i}}$, and $v_i$ is the perturbed three-velocity of the DM fluid, where $i = 1,2,3$ denotes the spatial index. It should be noted that Eq.~\eqref{evolution} and Eq.~\eqref{perturbations} are derived from the assumption that the DM is a pressureless perfect fluid, and thus its average density evolution is unaffected by the momentum coupling.

On the other hand, the Euler equation is modified by the momentum coupling, and it is derived from contracting Eq.~\eqref{Energy_Tensor} with the perturbed four-velocity $v_{\nu}$, as in \cite{palma2023}, 
\begin{eqnarray}
\nonumber \nabla^{\mu}(T^{(\phi)\nu}_{\mu} v_{\nu})   =  0 \to \\ 
  \dot{v}_i + \gamma_1{\cal{H}} v_i + \gamma_2 \partial_i \Psi 
 =& 0, \label{eq:euler eq} 
\end{eqnarray}
with 
\begin{eqnarray}
\label{coefficients}
\nonumber c_3 & = & \frac{2\beta\dot{\phi}^2}{a^2 \rho_c - 2 \beta \dot{\phi}^2}, \\
\nonumber c_1 & = & 2\beta c_3, \\
 c_2 & = & 
\nonumber\left[ \left(3- 4\beta\right) + 2 a^2  \frac{dV}{d\phi} \frac{1}{\cal{H} } \right]c_3,\\
\nonumber  \gamma_1 & = & \frac{1+c_2}{1+c_1}, \\
\gamma_2 & = & \frac{1+c_3}{1+c_1}. \label{eq: coefficients}
\end{eqnarray}
It is clear that the coupling strength $\beta$ affects the magnitude of the cosmological friction and the gravitational force.  In Eq.~\eqref{eq: coefficients}, because the first term of the denominator of $c_3$, $a^2 \rho_c$, is always positive, a large deviation from the uncoupled case ($c_{j} = 0,\,\, j = 1, 2, 3$) is expected when $a^2\rho_{c}  \approx 2\beta\dot{\phi}^2$. 

In addition, the quantitative effects of the coupling to the Euler equation can be further studied by focusing on the coefficients of the friction and gravitational force terms, namely $\gamma_1$ and $\gamma_2$ in Eq.~\eqref{eq: coefficients}, respectively. It is these two terms that encode the momentum transfer between DE and DM at the level of linear perturbations for all nonzero values of $ \beta $. When the momentum coupling is absent $\beta=0$, $c_{j}=0$, and $\gamma_{1,2} = 1$. The signs of $\gamma_{1,2}$ bear significant implications for structure formation. At positive values, $\gamma_{1}$ quantifies the extent to which the motion of the DM fluid is damped by the expansion of the Universe, since the cosmological friction is antiparallel to the velocity of the fluid. Therefore, an increase in $\gamma_{1}$, when positive, means that the fluid must overcome stronger friction to collapse to form structures. Conversely, for negative $\gamma_{1}$, the friction force facilitates the fluid's motion and enhances structure formation at linear scales. Likewise, $\gamma_{2}$ represents the effective strength and direction of the gravitational potential acting on the fluid, which, at positive values, also speeds up the formation of structures. As reported by \cite{palma2023}, if $\gamma_1 < 0$ and $\gamma_2 > 0$, the resulting collapsed halos exhibit greater velocity dispersion compared to the uncoupled case, and it is more difficult to achieve virial equilibrium while the magnitudes of $\gamma_{1,2}$ increase over time, leading to dynamical instability in the system.



\subsection{Implementations of simulations}
\label{sec:simulations}

The main features of the Type 3 model, a quintessence scalar field with momentum coupling, are encapsulated by $\beta$ and $\lambda$. Therefore, we focus on these two parameters to probe the relationship between the momentum transfer in the dark sector and the mean peculiar pairwise velocity. 

To study how these parameters affect the cosmology at the linear level, we performed Markov chain Monte Carlo (MCMC) analysis using \texttt{Cobaya} \cite{Torrado2021}. In \cite{Pourtsidou2025}, MCMC constraints on $\beta$ and $\lambda$ were obtained by refitting the Type 3 model with Planck CMB and lensing \cite{planck2020b, Rosenberg2022, Carron2022, Aghanim2019}, DESI DR2 BAO \cite{AbdulKarim2025}, and the DES-Y5 Type Ia supernova \cite{DESCollaboration2024}. Using the same datasets, we ran additional MCMC chains for different combinations of $\beta$ and $\lambda$ allowed by the previous constraints, which fall within the 1$\sigma$ deviations of the parameters reported in \cite{Pourtsidou2025} (see Appendix \ref{sec: mcmc constraints}). Our chains converge when the Gelman-Rubin diagnostic \cite{Gelman1992} $R-1 \leq 0.01$, and they are analyzed using \texttt{GetDist} \cite{Lewis2025}. This analysis is essential to the N-body simulation, as the cosmological parameters govern the initial conditions, the expansion history, and the clustering of matter. 



To incorporate the interaction between DM and DE, we modified the cosmological evolution accordingly. We used a modified version of \texttt{2LPTic} \cite{Crocce2006} to generate initial conditions corresponding to each combination of $\beta$ and $\lambda$, ensuring consistency with the input cosmological parameters. These initial conditions were created using the capacity constrained Voronoi tessellation (CCVT) method \cite{Liao2018, Zhang2021}, which produces a uniform and isotropic particle distribution. Compared to the gravitational equilibrium glass distribution \cite{White1996}, the CCVT distribution is better suited for models involving interactions beyond gravity and achieves geometric equilibrium. The initial matter power spectrum at redshift $z=99$ was then computed
from \texttt{CLASS} \cite{lesgourgues2011}, which was adapted for the Type 3 model.

Appendix~\ref{sec: Power spectrum} presents the linear matter power spectrum for both the $\Lambda$CDM model and the Type 3 model with zero and nonzero momentum exchange parameter $\beta$. Compared to the $\Lambda$CDM limit where $\beta = 0$ and $\lambda = 0$, Fig.~\ref{fig:power spectrum} illustrates how variations in $\beta$ and $\lambda$ affect the linear matter power spectrum. These deviations from the $\Lambda$CDM can be physically explained by the variations of these two parameters, allowing the Type 3 model to be clearly distinguished from $\Lambda$CDM. This robust result provides a strong foundation for subsequent N-body simulations.

We investigated ten combinations of $\beta$ and $\lambda$, derived from their allowed ranges constrained using the MCMC as presented in Table~\ref{tab:mcmc_cases}, along with their associated cosmological parameters. For the simulations, we utilized \texttt{ME-Gadget-2} \cite{An2019a, An2019b, zhang_gadget_2018}, a modified version of \texttt{Gadget-2} \cite{Springel2005}. This code handles modified cosmological evolution by using precomputed tables at discrete scale factors. In our work, we first simulated the evolution of linear perturbations using \texttt{CLASS}. From these simulations, we obtained and modified the tables that govern the Hubble expansion rate $H(a)$, and the coefficients of the friction and gravitational force, $\gamma_1(a)$ and $\gamma_2(a)$, respectively, as defined in Eq.~(\ref{coefficients}). The simulation set features a CDM particle count of $1024^3$, a grid resolution of $1024^3$, and a box size of $L_{\text{box}} = 1000 \, \text{Mpc} \, h^{-1}$, where $h$ is the dimensionless Hubble constant defined by $H_0 = h \times  100 \,  \, \text{km} \, \text{s}^{-1} \, \text{Mpc}^{-1}$.

DM halos are identified using \texttt{ROCKSTAR} \cite{Behroozi2012}, which employs the 3D friends-of-friends method to build a hierarchy of subgroups by analyzing the phase space of particles (3D position and 3D velocity) and reducing the 6D linking length to progressively uncover denser clusters of particles. Following \cite{Behroozi2012}, we define the halo center as the average position of particles in the inner subgroup of the halo, and the halo velocity as the average velocity of particles within the innermost 10\% of the halo's virial radius.

\begin{figure*}
    {\raggedright \hspace*{2.3em} \scalebox{1.2}{\textbf{\( \beta =-1.4, \,\,\lambda=0.6\)}} \raggedright \hspace*{6.1em} \scalebox{1.2}{\textbf{\( \beta =-0.5, \,\,\lambda=0.6 \)}} \raggedright \hspace*{6.1em} \scalebox{1.2}{\textbf{\( \beta =-0.5, \,\,\lambda=1.4 \)}} \par} 
    \includegraphics[width=0.95\textwidth]{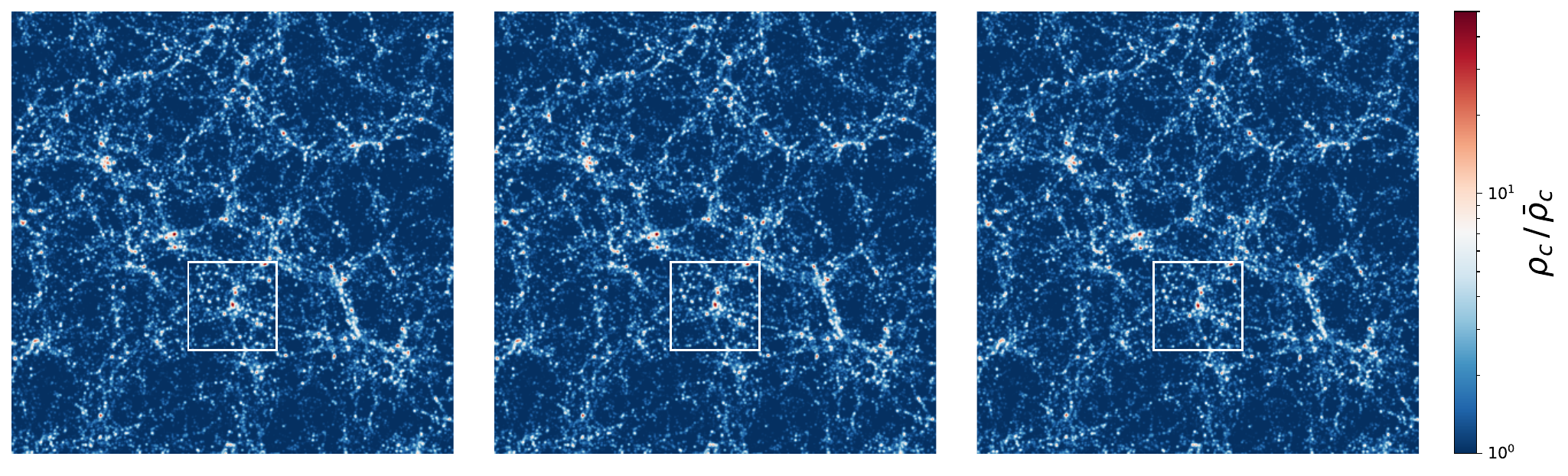}
    {\raggedright \hspace*{0.2em} \scalebox{1}{\textbf{\( 400 \times 400 
\times 80  \,\, \mathrm{Mpc}^{3} \,\, \mathrm{h}^{-3}\)}} \par}             
    \includegraphics[width=0.95\textwidth]{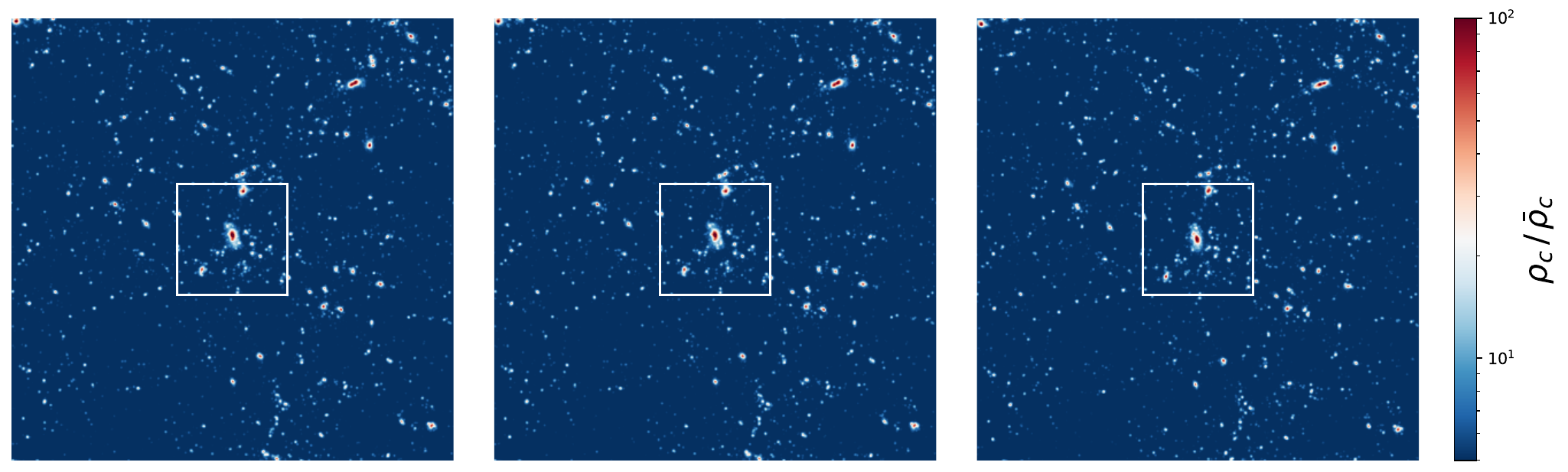}
    {\raggedright \hspace*{0.2em} \scalebox{1}{\textbf{\( 80 \times 80 
\times 80  \,\, \mathrm{Mpc}^{3} \,\, \mathrm{h}^{-3}\)}} \par}
    \includegraphics[width=0.95\textwidth]{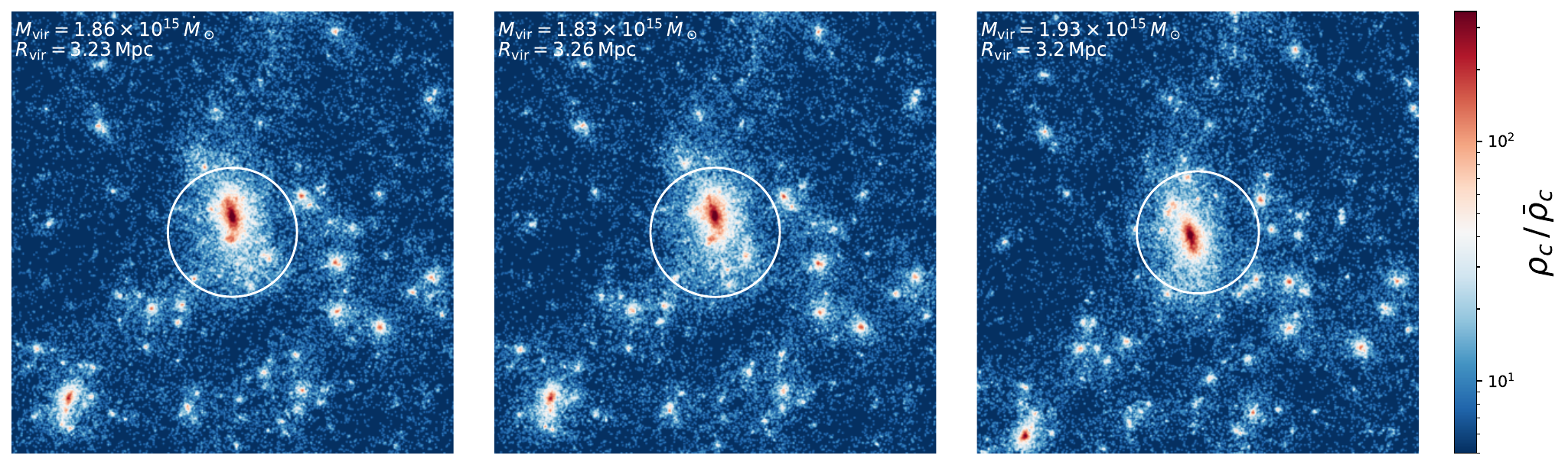}
    {\raggedright \hspace*{0.2em} \scalebox{1}{\textbf{\( 20 \times 20 
\times 60  \,\, \mathrm{Mpc}^{3} \,\, \mathrm{h}^{-3}\)}} \par}
\caption{2D overdensity distributions of CDM at redshift \( z = 0 \). The left, center, and right columns correspond to parameter configurations A1 ($\beta=-1.4$, $\lambda=0.6$), A7 ($\beta=-0.5$, $\lambda=0.6$), and A9 ($\beta=-0.5$, $\lambda=1.4$), respectively. Each row provides a progressively enlarged view of the regions of interest (white squares in top two rows) with the spatial scale and slice thickness indicated below the figures. The bottom row highlights a host halo enclosed by a white circle corresponding to its virial radius (\( R_{\text{vir}} \)). The corresponding virial mass (\( M_{\text{vir}} \)) and \( R_{\text{vir}} \) values are annotated in the upper-left corners. The density scale is indicated by the color bars. }
\label{fig:density_plots_T3}
\end{figure*}

Table~\ref{tab:mcmc_cases} presents the refitted parameters used in the N-body simulations for the ten cases, while Fig.~\ref{fig:density_plots_T3} shows the corresponding projected 2D overdensity distributions. In these figures, the DM host halo is encircled, highlighting three parameter configurations: A1 ($\beta = -1.4$, $\lambda = 0.6 $), A7 ($\beta = -0.5$, $\lambda = 0.6 $), and A9 ($\beta = -0.5$, $\lambda = 1.4 $). The LSS across the different cases appears broadly similar at first glance. However, a closer examination reveals a degeneracy between the effects of $\beta$ and $\lambda$ on the nonlinear structures. 

At the one-halo region, DM particles within host halos primarily exhibit orbital motion, as they are gravitationally bound and orbit the halo center. Based on Eq.~\eqref{scf_lag}, the momentum coupling between DM particles and the scalar field introduces an additional term in the scalar field Lagrangian, $\beta Z^2$. If $\beta < 0$, this term represents a transfer of momentum from the DM particles to the scalar field, introducing an additional frictional force to slow down the motion of DM particles compared to the uncoupled scenario. As discussed in \cite{baldi2015, palma2023}, for DM particles bound within the halo, this frictional effect facilitates the loss of angular momentum, enhancing their collapse toward the halo center. As a result, the virial equilibrium of the halo is altered, leading to its contraction relative to the uncoupled case.

This effect becomes more pronounced as the scalar field potential steepens, because the scalar field evolves faster ($\dot\phi$ increases), strengthening the kinetic coupling and the momentum interaction between the dark components. Consequently, DM particles experience greater friction, as evidenced by greater $\gamma_1$ values in Table~\ref{tab:cosmo_params}, leading to more efficient collapse and denser halo structures. Conversely, the momentum transfer process is hindered if the coupling is weaker, reducing the frictional damping ($\gamma_1$ decreases). This results in less efficient clustering at the one-halo region, leaving DM particles more dispersed and halos less compact compared to cases with stronger coupling \cite{baldi2015, palma2023}.

In this study, we explore the degeneracy between $\beta$ and $\lambda$ using the matter pairwise velocity statistics.

\begin{table*}
\large
\centering
\begin{tabular}{c c c c c c c c c c c}
\hline
\hline
No. & $\beta$ & $\lambda$ & $H_0\ (\mathrm{km\ s^{-1}\ Mpc^{-1}})$ & $\Omega_b$ & $\Omega_c$ & $\Omega_m$ & $n_s$ & $A_s\ (10^{-9})$ & $\gamma_1$ & $\gamma_2$ \\
\hline
A0 & 0 & 0.6 & 67.21 & 0.0494 & 0.2602 & 0.3097 & 0.9679 & 2.115 & 1 & 1 \\
\hline
A1 & $-$1.4 & 0.6 & 67.78 & 0.0485 & 0.2568 & 0.3052 & 0.9670 & 2.111 & 1.4281 & 0.9021 \\
A2 & $-$1.4 & 1 & 67.42 & 0.0490 & 0.2588 & 0.3081 & 0.9676 & 2.116 & 1.9779 &  0.7711 \\
A3 & $-$1.4 & 1.4 & 66.86 & 0.0499 & 0.2623 & 0.3126 & 0.9685 & 2.122 & 2.4931 & 0.6379 \\
\hline
A4 & $-$0.8 & 0.6 & 67.68 & 0.0487 & 0.2574 & 0.3061 & 0.9671 & 2.112 & 1.3597 &  0.9173 \\
A5 & $-$0.8 & 1 & 67.15 & 0.0495 & 0.2607 & 0.3103 & 0.9681 & 2.117 & 1.8266 & 0.8032 \\
A6 & $-$0.8 & 1.4 & 66.34 & 0.0508 & 0.2659 & 0.3169 & 0.9693 & 2.125 & 2.2611 & 0.6834 \\
\hline
A7 & $-$0.5 & 0.6 & 67.60 & 0.0488 & 0.2578 & 0.3066 & 0.9673 & 2.114 & 1.2943 & 0.9319 \\
A8 & $-$0.5 & 1 & 66.90& 0.0499 & 0.2657 & 0.3123 & 0.9684 & 2.120 & 1.6798 & 0.8355 \\
A9 & $-$0.5 & 1.4 & 65.83 & 0.0516 & 0.2697 & 0.3214 & 0.9699 & 2.129 & 2.0353 & 0.7312\\
\hline
\hline
\end{tabular}
\caption{Refitted parameters employed in the simulations. $\beta$ is the momentum coupling constant, and $\lambda$ is the scalar field potential slope. $H_0$ is the Hubble constant, and $\Omega_{x}$ represents the energy density parameters, with the index $x\in\{b,c,m\}$. The spectral index $n_s$ and the amplitude $A_s$ quantify the properties of primordial scalar perturbations. Lastly, $\gamma_1$ and $\gamma_2$ are the friction and modified gravitational coefficients at $z=0$, respectively. }
\label{tab:mcmc_cases}
\end{table*}

\section{Pairwise velocity statistics} \label{sec:velocity}
    \subsection{Pairwise velocity}
    For a particle pair at separation $r$ in the comoving frame, the mean peculiar pairwise velocity is given by 
\begin{equation}
        v_{12}(r) =  \left<  \left( \mathbf{v}_1(\mathbf{r}_1) - \mathbf{v}_2(\mathbf{r}_2) \right) \cdot \hat{\mathbf{r}} \right>.
\label{eq: v12}
\end{equation}

Here, $\mathbf{v}_1$ and $\mathbf{v}_2$ are the peculiar velocities of particles 1 and 2, $\mathbf{r}_1$ and $\mathbf{r}_2$ their respective comoving positions, and $\hat{\mathbf{r}} = (\mathbf{r}_2 - \mathbf{r}_1)/|\mathbf{r}_2 - \mathbf{r}_1|$ is the unit separation vector between the pair. The angle brackets $\langle \cdots \rangle$ denote the ensemble average over all particle pairs separated by a comoving distance $r$.  

The origin of $ v_{12} $ arises from the gravitational interactions between particles. This reflects a tendency for particles to move toward each other when cosmological expansion is excluded. At small separations, the particles become gravitationally bound and clustered. This means that their mean physical relative velocity $ v^{phy}_{12} \approx 0 $. As $r$ increases, gravitational force weakens, causing the particles' peculiar motions to become uncorrelated, and their mean peculiar pairwise velocity $ v_{12} \to 0 $. The mean peculiar pairwise velocity is given by \cite{baumann2022}
\begin{equation}
    v_{12}(r) = v^{phy}_{12} - Hr^{phy}, \label{velocity}
\end{equation}
where $H$ is the Hubble parameter and $r^{phy} = a(t)r$.

From the conservation of the total particle number, the average rate of change in the number of particles ${N(r,t)}$ within a sphere of comoving radius $ r $ centered on a target particle is \cite{Mo2010}
\begin{equation}
\frac{\partial N(r,t)}{\partial t}  = -4\pi \overline{n} a^2 r^2 \left[ 1 + \xi(r,a) \right] v_{12}(r,a), \label{number}
\end{equation}

where $\overline{n}$ denotes the average comoving number density of particles over the spherical shell at time $t$, and $\xi(r,a)$ is the two-point correlation function. The right-hand side of Eq.~\eqref{number} represents the average flux of the particles per unit time (over the surface of the spherical shell) passing into the spherical shell of radius $r$ that is centered on the target particle.

By combining Eqs.~\eqref{velocity}-\eqref{number}, and using the conservation of particle number in an expanding Universe, $\bar{n}(t)a^3 = constant$, we have
\begin{equation}
\frac{v_{12}(r,a)}{Hra} = -\frac{a}{3(1+\xi)}\frac{\partial \bar{\xi}(r,a)}{\partial a}.  \label{vel_two-point}
\end{equation}
Here, the average two-point correlation function is defined as
\begin{equation}
\bar{\xi}(r, a) = \frac{3}{4\pi r^3} \int d^3\mathbf{y} \, \xi(y, a).  
\end{equation}

When $r$ is large, $\xi(r,a) \to 0$. Therefore, one can expand $\xi(r, a)$ to the $1^{\text{st}}$ order in density perturbations as  $\xi(r,a) = \xi(r,a=1)D^2 (a)$, obtaining the linear approximation of the mean peculiar pairwise velocity \cite{Juszkiewicz1999, Peebles2020}
\begin{equation}
v_{12}(r,a) = -\frac{2}{3}Hra f(a) \bar{\bar{\xi}}(r,a),
\label{linear_velocity}
\end{equation}

with $D$ being the linear growth factor that arises from the linear approximation to ${\xi}(r,a)$ and $f(a) = \dv{\ln D}{\ln a} $ is empirically approximated as $f(a) \equiv \Omega_{m}^{0.55}(a)$, where $\Omega_{m}$ represents the total matter energy density parameter \cite{Linder2005}. The normalized correlation function $\bar{\bar{\xi}}(r, a)$ is given by  $  \bar{\bar{\xi}}(r, a) = \frac{\bar{\xi}(r, a)}{1 + \xi(r, a)}$. Eq.~\eqref{linear_velocity} illustrates that the mean peculiar pairwise velocity $v_{1,2}(r, a)$ encapsulates both spatial clustering information [via $\xi(r, a)$] and its temporal evolution [via $f(a)$)]. 

\subsection{Pairwise velocity dispersion}

The mean peculiar pairwise velocity dispersion is given by 
\begin{equation}
        \sigma_{12}(r) =  \langle \{  \left[ \mathbf{v}_1(\mathbf{r}_1) - \mathbf{v}_2(\mathbf{r}_2) \right] \cdot \hat{\mathbf{r}} \}^{2}\rangle^{1/2}  .
\label{eq: sigma12}
\end{equation}

As demonstrated in Appendix \ref{sec: BBGYK equations}, integrating the momenta of the second Bogoliubov–Born–Green–Kirkwood–Yvon (BBGKY) equation yields the cosmic virial theorem (CVT) under the condition ($\zeta \gg \xi \gg 1$) \cite{Mo2010, Peebles2020, Mo1997}, in which $\zeta$ is the
three-point correlation function. This theorem enables the approximation of $\sigma_{12}^{2}(r)$ at various separation scales, giving

\begin{equation}
\sigma_{12}^2(r) =
\begin{cases}
\frac{3\Omega_m(a) H^2(a)}{4\pi \xi(r,a)} \int_r^\infty \frac{dr'}{r'} \int d^{3}z \, \frac{\mathbf{r'}\cdot\mathbf{z}}{z^{3}} \zeta(r', z, |\mathbf{r'} - \mathbf{z}|, a) \\
\text{ at small } r \\
\\
\frac{2}{3}\langle v^2 \rangle \\
\text{ at large }  r . \\
\end{cases}  \label{CVT}
\end{equation}

Similarly, by integrating over the momenta of the first BBGKY equation, the cosmic energy equation (CEE) can be obtained, and its subsequent integration leads to an estimation for $\langle v^2 \rangle$, 

\begin{equation}
\langle v^2 \rangle  = \frac{3}{2} \Omega_m(a) H^2(a) a^2 {\cal{I}}_2(a) \left[ 1 - \frac{1}{a {\cal{I}}_2(a)} \int_0^a da^{'} {\cal{I}}_2(a^{'}) \right], \label{CEE_v}
\end{equation}
where ${\cal{I}}_2(a) \equiv \int_0^\infty dr \, \xi(r, a) r. $

It can be seen that both the CVT and CEE stem from the conservation of Newtonian energy and that they capture the connections between the kinetic energy in the form of $\langle v^2 \rangle $ and the mean gravitational potential energy of the system originating from all particle pairs.

In \cite{zhang2024}, the effects of neutrino mass $M_v$ and asymmetry $\eta^{2}$ on the matter peculiar pairwise velocity and dispersion were extensively studied. In this paper, we extend the investigation to examine the dependence of pairwise velocity statistics
 on the momentum coupling constant $\beta$ and the scalar field potential slope $\lambda$ of the Type 3 model.


 \section{Results} \label{sec:results}

This Section presents the results for \( v_{12}(r) \) and \( \sigma_{12}(r) \), both evaluated at \( z = 0 \), from simulations with a box size of \( L_{\text{box}} = 1000 \, \text{Mpc} \, h^{-1} \). The scalar field and its coupling to DM introduce three distinct effects compared to $\Lambda$CDM: a modified expansion history of the Universe, an additional cosmological friction force and reduced gravity acting on DM, and a set of modified cosmological parameters by refitting the aforementioned datasets. According to \cite{chamings2020}, a faster expansion rate mildly suppresses structure growth by reducing the magnitude of the metric perturbation term \( |\dot{\Phi}| \) in Eq.~\eqref{perturbations}. In contrast, the altered cosmological friction and the refitted cosmological parameters are more consequential to the structure formation.

\subsection{Particle-particle analysis}

As demonstrated in Appendix~\ref{sec: CDM fraction}, it is sufficient to randomly sample $1\%$ of CDM particles for calculating the mean particle-particle peculiar pairwise velocity ($v_{\mathrm{pp}}$) and velocity dispersion ($\sigma_{\mathrm{pp}}$), because the difference in the measurements with $10\%$ and $1\%$ sampling rates is negligible. For this reason, our results are primarily based on a $1\%$ random sample of CDM particles. 

Furthermore, given our simulation resolutions, the uncertainty in the measurement of pairwise velocity statistics due to cosmic variance can be neglected, as demonstrated in Appendix\ref{sec: Cosmic variance}. Using different random seeds to generate the initial conditions (ICs), the mean deviations in $v_{\mathrm{pp}}$ and $\sigma_{\mathrm{pp}}$ from the uncoupled scenario (A0) for various $\beta$ values may fluctuate slightly, but they gradually converge at larger scales. 

\begin{figure*}
    \includegraphics[width=.95\textwidth]{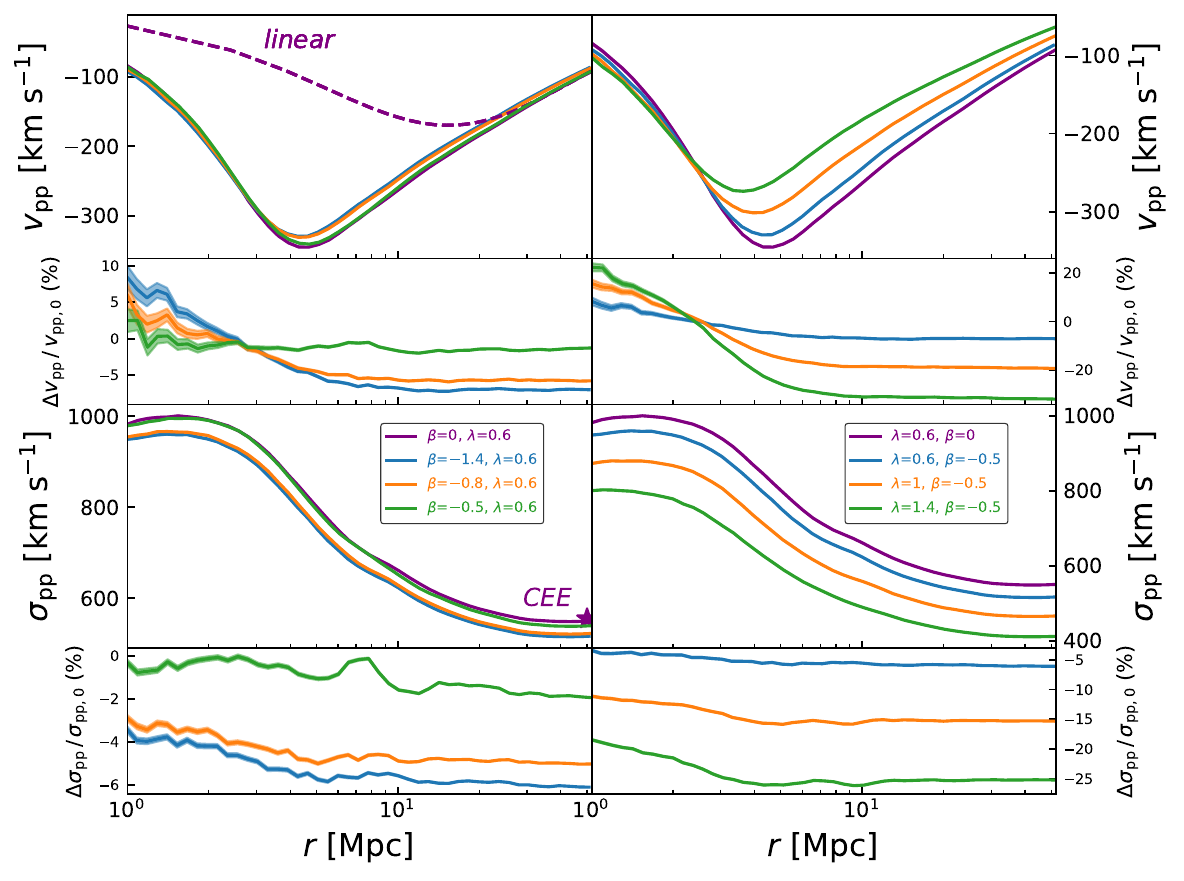}
    \caption{Mean particle-particle peculiar pairwise velocity $v_{\mathrm{pp}}$ (top panel) and velocity dispersion $\sigma_{\mathrm{pp}}$ (bottom panel). The left and right columns show the pairwise velocity statistics
 for varying $\beta$ (with $\lambda=0.6$) and $\lambda$ (with $\beta=-0.5$), respectively, compared to the uncoupled case A0 with $\beta = 0, \, \lambda = 0.6$ (purple lines). Subpanels display percentage deviations from the fiducial configuration's (A0), $\frac{\Delta [v/\sigma]_{\mathrm{pp}}}{[v/\sigma]_{\mathrm{pp,0}}}= \frac{[v/\sigma]_{\mathrm{pp}} (\text{other cases}) - [v/\sigma]_{\mathrm{pp}} (\mathrm{A0})}{[v/\sigma]_{\mathrm{pp}} (\mathrm{A0})}$. The dashed line in the top left panel represents the linear approximation, and the star in the bottom panel denotes the CEE approximation, both for A0.}\label{fig:pp-vel}
\end{figure*}

Fig.~\ref{fig:pp-vel} shows the $v_{\mathrm{pp}}$ ($\sigma_{\mathrm{pp}}$) in the top (bottom) row. The lower subpanels illustrate the percentage deviations from the fiducial scenario (A0) with no momentum coupling ($\beta=0$). The left column demonstrates the impact of varying $\beta$ for fixed $\lambda=0.6$ on the pairwise velocity statistics, while the right column shows the statistics for varying $\lambda$ at fixed $\beta=-1.4$. 

The dashed line in the top subpanel of the left column shows the linear approximation of $v_{\mathrm{pp}}$ for A0 based on Eq.~\eqref{linear_velocity}. Here, $\xi$ is calculated from the inverse Fourier transform (IFT) of the linear matter power spectrum at $z=0$ obtained from \texttt{CLASS}. At large separations $r$, the linear approximation closely matches the measurement from our simulations. However, when entering into the nonlinear regime, which covers the majority of the range of $r$, the linear approximation becomes inadequate and severely underestimates the actual measurement. 

Similarly, the star in the lower panel of the left column denotes the linear approximation of $\sigma_{\mathrm{pp}}$ for A0 derived from the estimation of $\langle v^2 \rangle$ in the CEE, Eq.~\eqref{CEE_v}. Unlike the $v_{\mathrm{pp}}$ approximation, $\xi$ within ${\cal{I}}_2$ is the IFT of the nonlinear halofit matter power spectrum, in which $k$ is attenuated below $2\pi/L_{box}$ to account for the finite box effect \cite{Bird2012}. It can be seen that the approximation is in close agreement with the measurement. 

While the linear approximations of $v_{\mathrm{pp}}$ and $\sigma_{\mathrm{pp}}$ agree with our simulation results for A0 in the large $r$ limit, it should be noted that neither the linear growth rate function $f(a)$ in Eq.~\eqref{linear_velocity} nor the halofit function has taken into account the coupling effect within the dark sector. Therefore, the approximations are only presented for the uncoupled scenario  $\beta=0$ (A0).   

Fig.~\ref{fig:pp-vel} shows that the minimum of  $v_{\mathrm{pp}}$ occurs at  $r \approx 4 \,\text{Mpc}$, corresponding to the intermediate scale between the one-halo and two-halo regime. As $r \to 0 $ or $\infty$, $v_{\mathrm{pp}}$ stabilizes and approaches $0$. This behavior is consistent with theory: at small interparticle separations, DM particles form clusters, while at large separations, they exhibit stochastic, uncorrelated motions.

As shown in the top panel of Fig.~\ref{fig:pp-vel}, for $r < 3 \, \mathrm{Mpc}$, all cases exhibit a greater magnitude of $v_{\mathrm{pp}}$ than the uncoupled case (purple line) with $\lambda=0.6$, corresponding to friction ($\gamma_1$) and gravitational ($\gamma_2$) coefficients of 1. This is expected because, at nonlinear scales, DM particles are gravitationally bound to host halos. The additional friction exerted on these particles---due to momentum coupling with the scalar field---accelerates the loss of their angular momentum, as evidenced by the increased friction coefficients $\gamma_1 > 1$ of all cases relative to the uncoupled case. This process facilitates their collapse into the central regions of halos, thereby increasing $|v_{\mathrm{pp}}|$. Specifically, at fixed $\lambda = 0.6 $, the left panel shows that a smaller $|\beta|$ (weaker momentum coupling) corresponds to lower $|v_{\mathrm{pp}}|$. In contrast, for fixed $\beta =-1.4$, the right panel shows that a larger $\lambda$ (steeper potential slope) would increase $\left|v_{\mathrm{pp}}\right|$, and this reflects the idea that a steeper potential drives the evolution of the scalar field $\dot\phi$, enhancing momentum transfer within the dark sector. This strengthens the frictional force on DM particles (larger $\gamma_1$), facilitating angular momentum loss and driving particles toward halo centers. 


At larger separations ($r > 3 \, \mathrm{Mpc}$), the trend reverses: all cases show suppressed $|v_{\mathrm{pp}}|$ relative to the uncoupled case. In this linear regime, structure formation is dominated by radial infall, with DM particles exhibiting primarily linear motion. The transition scale ($r \sim 3 \, \mathrm{Mpc}$) corresponds to the shell-crossing radius, which marks the boundary between the nonlinear regime (collapsed structures with angular dynamics) and the linear regime (growing perturbations dominated by radial flows). Shell-crossing occurs when initially distinct spherical shells of matter around an overdensity cross each other due to gravitational collapse, as outer shells collapse faster and overtake inner ones, leading to multistream flows and the onset of virialized motion within halos. In the linear regime, weaker coupling (smaller $|\beta|$) reduces the frictional damping of radial motion, allowing faster infall and more efficient clustering, which increases $|v_{\mathrm{pp}}|$ compared to strongly coupled cases. In contrast, a steeper potential (larger $\lambda$) amplifies friction (larger $\gamma_1 >1$) and weakens  gravitational interactions between particle pairs (lower $\gamma_2$), leading to slower radial infall and clustering. This reduced efficiency of structure formation further lowers $|v_{\mathrm{pp}}|$. 

Unlike $v_{\text{pp}}$, the effects of $\beta$ and $\lambda$ on $\sigma_{\text{pp}}$ are consistent across all scales studied and follow the trends of $v_{\text{pp}}$ in the linear regime. This difference stems from the distinct definitions of $v_{\text{pp}}$ and $\sigma_{\text{pp}}$: in the expression for $\sigma_{12}^2(r)$ in Eq.~\eqref{eq: sigma12 los}, $\frac{2}{3} \langle v^2 \rangle$ represents the average kinetic energy of a DM particle pair. As $|\beta|$ or $|\lambda|$ decreases, the average kinetic energy of the particle pair increases, as reflected by a decreased friction coefficient $\gamma_1$ in Table~\ref{tab:mcmc_cases}. This leads to an increased velocity dispersion $\sigma_{\text{pp}}$, matching the trend of $v_{\text{pp}}$ in the linear regime. Furthermore, the second term, $\Pi$, reflects the correlated motion of the pair parallel to their line of separation, making it primarily sensitive to radial motion, which dominates in the linear regime. Therefore, $\sigma_{\mathrm{pp}}$ is sensitive primarily to radial motion, and it depends on overall kinetic energy and radial correlations, both of which increase when friction decreases.

\begin{figure*}
    \includegraphics[width=.95\textwidth]{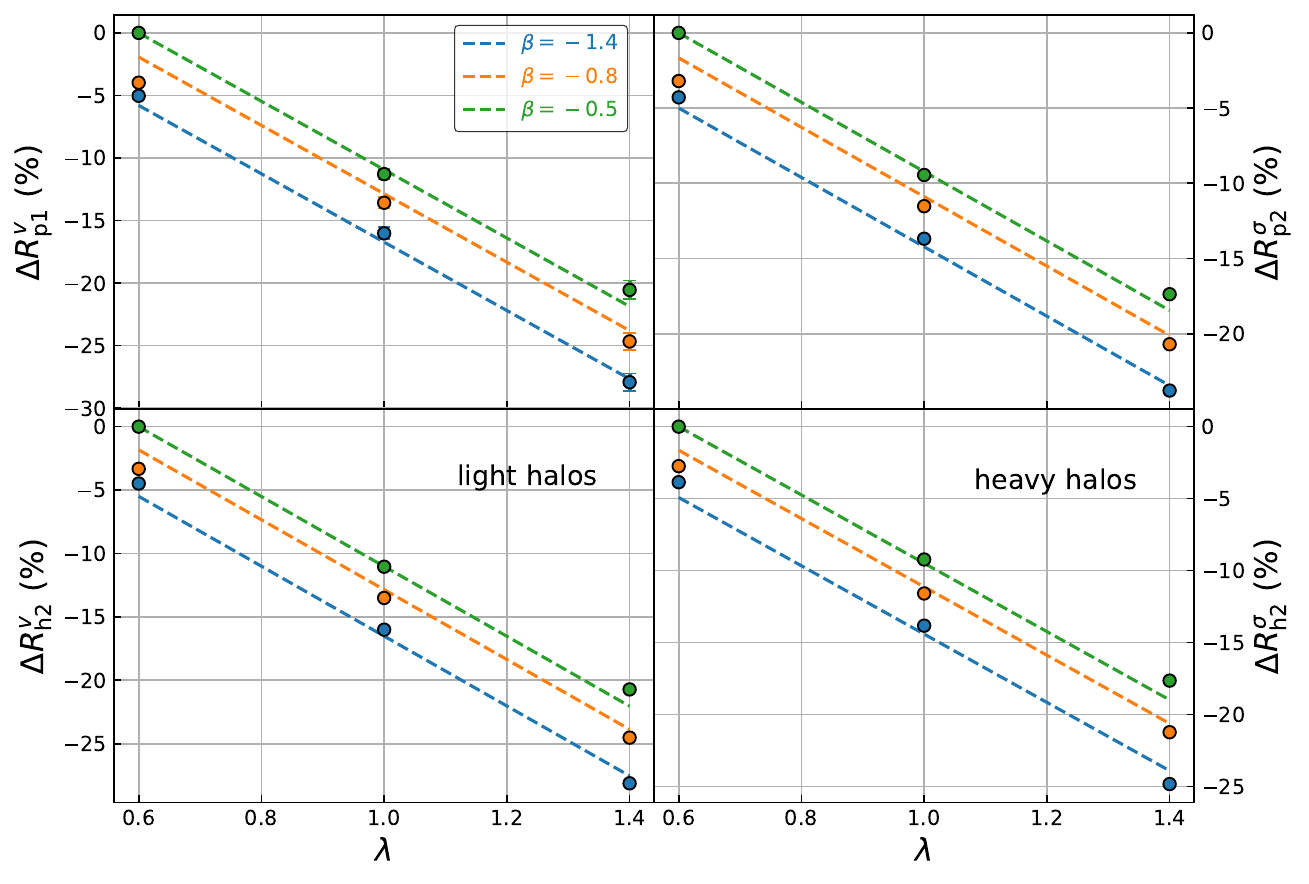}
    \caption{$\Delta R^{v}_{\text{p1}}$ (upper left), $\Delta R^{\sigma}_{\text{p2}}$(upper right), $\Delta R^{v}_{\text{h2}}$ for light halos (lower left), and $\Delta R^{\sigma}_{\text{h2}}$ for heavy halos  (lower right), respectively.  Displayed as a function of $\lambda$, the data obtained from the simulations, along with their error bars, are fitted to a three-variable linear regression (dashed lines), with different colors corresponding to different $\beta$.}  \label{fig:deviations-vel}
\end{figure*}
Similar to \cite{zhang2024}, to summarize the simulation results, we performed a quantitative analysis of $v_{\mathrm{pp}}$ and $\sigma_{\mathrm{pp}}$ by studying their ratios $R^{v/\sigma}_{\mathrm{pp}}$, respectively, which are defined as 

\begin{equation}
R^{v/\sigma}_{\mathrm{pp}}(\beta,\lambda,r)=\frac{[v/\sigma]_{\mathrm{pp}}(\beta,\lambda,r)}{[v/\sigma]_{\mathrm{pp}}(-0.5,0.6,r)} \label{velocity_ratio}.
\end{equation}

Here, we adopt A7 (see Table~\ref{tab:mcmc_cases}) as the fiducial scenario instead of A0, in which  $[v/\sigma]_{\mathrm{pp}}(-0.5,0.6,r)$ correspond to its velocity statistics measurements. In addition, the average deviation from the A7 measurements over the radial interval $I_{k}$ is defined as 

\begin{equation}
    \Delta R^{v/\sigma}_{\mathrm{p}k}(\beta,\lambda) 
    = 
    \left.\overline{ R^{v/\sigma}_{\mathrm{pp}}(\beta,\lambda,r) }\right\vert_{r\in I_k} - 1,
\label{ratio}
\end{equation}

where $k$ stands for different $r$ intervals. In our study, we focus our analysis on the transition and two-halo regions that span between the following intervals: $I_{1} = [4,15] \,\text{Mpc} \,(k=1)$ and $I_{2} = [25,50] \,\text{Mpc} \,(k=2)$, respectively.


To quantify the relationship of the variables with $v_{\mathrm{pp}}$ and $\sigma_{\mathrm{pp}}$, we fit the measurements using the three-variable regression on $ \Delta R^{v/\sigma}_{\mathrm{p}k}(\beta,\lambda)$

\begin{equation}\label{eq:pp-pwv-fitting-formula}
    \Delta R^{v/\sigma}_{\mathrm{p}k}(\beta,\lambda)
    = C^{v/\sigma}_{\beta\mathrm{p}k}\Delta\beta+ C^{v/\sigma}_{\lambda\mathrm{p}k} \Delta\lambda .
\end{equation}

Here, $C^{v/\sigma}_{\beta\mathrm{p}k}$ and $C^{v/\sigma}_{\lambda\mathrm{p}k}$ denote the coefficients associated with the fitting parameters. The quantities $\Delta\beta = \beta + 0.5$ and $\Delta\lambda = \lambda - 0.6$ represent the deviations of the momentum coupling and potential slope parameters from the reference values in case A7 ($\beta = -0.5$, $\lambda = 0.6$), respectively.

Using Eq.~\eqref{eq:pp-pwv-fitting-formula} to fit our simulation results, we found the best-fit coefficients as displayed in Table~\ref{tab:fitting-results-pp}, and the fittings are illustrated in Fig.~\ref{fig:deviations-vel}. 

\begin{table*}
\centering
\begin{tabular}{cccc}
\hline
\hline
Range (Mpc) & Quantity & $C^{v/\sigma}_{\beta\mathrm{pk}}$ (\%) & $C^{v/\sigma}_{\lambda\mathrm{pk}}$ (\%) \\
\hline
{[4,15] } & $v_{\mathrm{pp}}$ & $6.44\pm0.81$ & $-27.33\pm0.86 $ \\
{$k=1$ } & $\sigma_{\mathrm{pp}}$ & $5.77\pm0.72$ & $-24.79\pm0.77$ \\
\hline
{[25,50]} & $v_{\mathrm{pp}}$ & $6.81\pm0.80$ & $-30.57\pm0.85$ \\
 {$k=2$ } & $\sigma_{\mathrm{pp}}$ & $5.55\pm0.64$ & $-23.06\pm0.68$ \\
\hline
\end{tabular}
\caption{Linear regression fitting results of $\Delta R^v_{\mathrm{p}k}$ and $\Delta R^\sigma_{\mathrm{p}k}$.}
\label{tab:fitting-results-pp}
\end{table*}

For $v_{\mathrm{pp}}$, the best-fit coefficients in the interval $r \in [4, 15] \,\text{Mpc}$ are  $C^{v}_{\beta \mathrm{p}1} (\%) =6.44  \, \pm 0.81$  and $C^{v}_{\lambda \mathrm{p}1} (\%)=-27.33 \,\pm 0.86$. These values show that, for the same direction of variation and near their fiducial values, the effects of $\beta$ ($<0$) and $\lambda$ act in opposite directions. This remains true in the interval $r \in [25, 50] \,\mathrm{Mpc}$, except that the effect of $\lambda$ on $v_{12}$ is slightly enhanced, as evidenced by the increase in the magnitude of the coefficient $C^{v}_{\lambda \mathrm{p}2} (\%)=-30.57 \,\pm 0.85$.  


For $\sigma_{\mathrm{pp}}$, in both the intermediate region $r \in [4, 15] \,\text{Mpc}$ and the two-halo region $r \in [25, 50] \,\text{Mpc}$, the same direction of variation in $\beta$ ($<0$) and $\lambda$ also act in opposite directions, with the effect of the latter overcoming the former.



The fittings for  $\Delta R^v_{\mathrm{p}k}$  and  $\Delta R^\sigma_{\mathrm{p}k}$ are illustrated in Fig.~\ref{fig:deviations-vel}.

\subsection{Halo-halo analysis}
Considering the broad distribution of halo masses, we categorize host halos into two virial mass ranges: light halos $[10^{11},10^{13}]\,M_{\bigodot}$, and heavy halos $[10^{13},10^{15}]\,M_{\bigodot}$. This choice of selection means that there are at least 200 CDM particles inside each host halo.

\begin{figure*}
    \includegraphics[width=.95\textwidth]{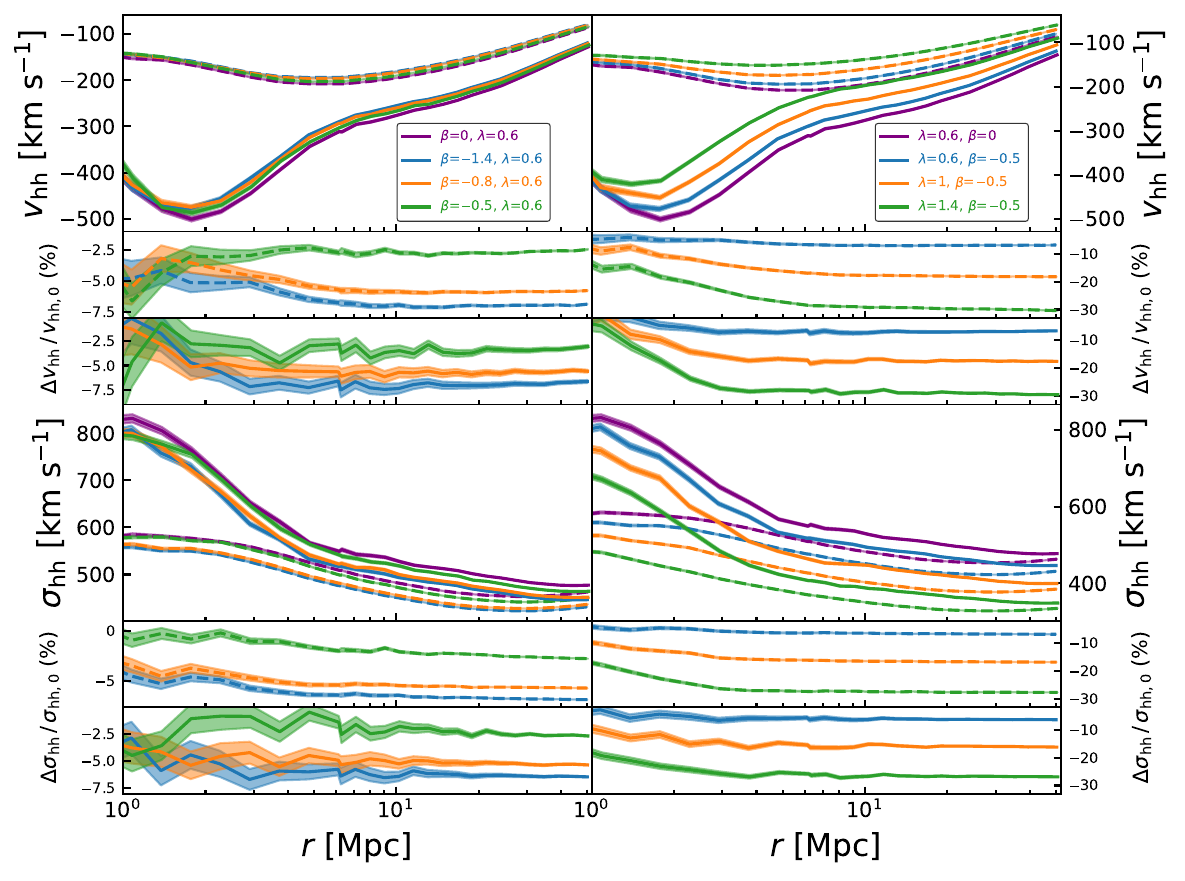}
    \caption{Mean halo-halo peculiar pairwise velocity $v_{\mathrm{hh}}$ (top panels) and velocity dispersion $\sigma_{\mathrm{hh}}$ (bottom panels), for light (dashed lines) and heavy (solid lines) halos. The left and right columns show the pairwise velocity statistics
 for varying $\beta$ (with $\lambda=0.6$) and $\lambda$ (with $\beta=-0.5$), respectively, compared to the uncoupled case A0 with $\beta = 0$, $\lambda = 0.6$ (purple lines). Subpanels display percentage deviations from the fiducial's, with $\frac{\Delta [v/\sigma]_{\mathrm{hh}}}{[v/\sigma]_{\mathrm{hh,0}}}= \frac{[v/\sigma]_{\mathrm{hh}} (\text{other cases}) - [v/\sigma]_{\mathrm{hh}} (\mathrm{A0})}{[v/\sigma]_{\mathrm{hh}} (\mathrm{A0})}$. }\label{fig:hh-vel}
\end{figure*}

Fig.~\ref{fig:hh-vel} shows that the magnitudes of $v_{\mathrm{hh}}$ and $\sigma_{\mathrm{hh}}$ of heavy halos are significantly greater than those for light halos. This discrepancy can be attributed to the fact that heavier halos reside in denser environments (halo bias), leading to a stronger gravitational pull and increased mean peculiar pairwise velocities. Additionally, the gravitational attraction between heavier halos is enhanced compared to lighter halos.

Another key distinction between light and heavy halos lies in the location of the minimum points $v_{\mathrm{hh}}$. For light halos, this occurs at $r \approx 8 \,\text{Mpc}$, which corresponds to the transition from the one-halo to the two-halo regimes. This is similar to the trend of $v_{\mathrm{pp}}$, but the transition separation for the halo-halo scenario is larger than that of the particle-particle scenario ($r \approx 4 \, \text{Mpc}$). For heavy halos, $v_{\mathrm{hh}}$ troughs at $r \approx 2 \,\text{Mpc}$, slightly exceeding the average virial radius of $1 \,\text{Mpc}$. Physically, this separation coincides with the splashback radius, where mergers dominate. Similar to the light halos, heavy halos also exhibit a transition at $r \approx 7 \,\text{Mpc}$, reflecting the shift from one-halo to two-halo regions. Beyond this separation, $v_{\mathrm{hh}}$ and $\sigma_{\mathrm{hh}}$ resemble those for the light halos but with greater magnitudes.


As shown in Fig.~\ref{fig:hh-vel}, across both intervals, for light halos (dashed lines), increasing $\beta$ consistently increases the magnitudes of $v_{\mathrm{hh}}$ and $\sigma_{\mathrm{hh}}$, while increasing $\lambda$ lowers the magnitudes. This indicates that the directionality and extent of the effects of these parameters on $v_{\mathrm{hh}}$ and $\sigma_{\mathrm{hh}}$ agree with the trends observed in $v_{\mathrm{pp}}$ and $\sigma_{\mathrm{pp}}$. Similarly, for heavy halos, the statistics remains sensitive to $\lambda$ and $\beta$ in both intervals. Despite the consistency in the trends between the $v_{\mathrm{hh}}$ and $v_{\mathrm{pp}}$, the former measurement does not capture the nonlinear effects of DM particles, where shell crossing is not observed at the boundary between the one-halo and intermediate regions. This is because $v_{\mathrm{hh}}$ measures the relative motion between halos, but not the internal motions of DM particles within halos.


Similar to the previous analysis for the particle-particle interaction, we study the ratios of the mean halo-halo velocity peculiar pairwise statistics to their average deviations from those of the case (A7),
\begin{equation}
\begin{split}
R^{v/\sigma}_{\mathrm{hh}}(\beta,\lambda,r) &= \frac{[v/\sigma]_{\mathrm{hh}}(\beta,\lambda,r)}{[v/\sigma]_{\mathrm{hh}}(-0.5,0.6,r)}, \\[2ex]  
\Delta R^{v/\sigma}_{\mathrm{h}k}(\beta,\lambda) &\equiv \left.\overline{ R^{v/\sigma}_{\mathrm{hh}}(\beta,\lambda,r) }\right\vert_{r\in I_k} - 1.
\end{split}
\label{combined_equation}
\end{equation}

Here, $k=1,2$ denotes the intervals $I_1=[4,15]\,\text{Mpc}$ and $I_2=[25,50]\,\text{Mpc}$, respectively. For light and heavy halos, we analyze both intervals to study the influence of $\beta$ and $\lambda$ in the intermediate and two-halo cluster regions. 


Similar to the particle-particle analysis, the average deviations derived from halos of different mass ranges are subsequently fitted to a two-variable linear regression, $ \Delta R^{v/\sigma}_{\mathrm{h}k}(\beta,\lambda)
    = C^{v/\sigma}_{\beta\mathrm{h}k}\Delta\beta+  C^{v/\sigma}_{\lambda\mathrm{h}k} \Delta\lambda .$ The fitting results of $\Delta R^{v}_{\mathrm{h}2}$ (lower left) and $\Delta R^{\sigma}_{\mathrm{h}2}$ (lower right)  are displayed in Fig.~\ref{fig:deviations-vel} for light and heavy halos, respectively. The best-fit coefficients are presented in Table~\ref{tab:fitting-results-hh}.

\begin{table*}
\centering
  \begin{tabular}{ccccc} 
    \hline
    \hline
    $M_{\mathrm{vir}}\ [M_\odot]$ & range [Mpc] & quantity
    & $C^{v/\sigma}_{\beta\mathrm{h}k}$ (\%) & $C^{v/\sigma}_{\lambda\mathrm{h}k}$ (\%) \\ 
    \hline
    \multirow{4}{*}{$[10^{11},10^{13}]$}
    & [4,15]
    & \multicolumn{1}{c}{$v_{\mathrm{hh}}$} & \multicolumn{1}{c}{$6.00\pm0.73$} & \multicolumn{1}{c}{$-25.79 \pm0.77$} \\ 
    & $k=1$ & \multicolumn{1}{c}{$\sigma_{\mathrm{hh}}$} & \multicolumn{1}{c}{$5.86\pm0.70$} & \multicolumn{1}{c}{$-25.08\pm0.74$} \\ 
    & {[25,50]}
    & \multicolumn{1}{c}{$v_{\mathrm{hh}}$} & \multicolumn{1}{c}{$6.11\pm0.71$} & \multicolumn{1}{c}{$-27.55\pm0.75$} \\ 
    & $k=2$ & \multicolumn{1}{c}{$\sigma_{\mathrm{hh}}$} & \multicolumn{1}{c}{$5.85\pm0.69$} & \multicolumn{1}{c}{$-24.53\pm0.74$} \\ 
    \hline
    \multirow{4}{*}{$[10^{13},10^{15}]$}
    & [4,15]
    & \multicolumn{1}{c}{$v_{\mathrm{hh}}$} & \multicolumn{1}{c}{$5.65\pm0.72$} & \multicolumn{1}{c}{$-24.61 \pm0.76$} \\ 
    & $k=1$ & \multicolumn{1}{c}{$\sigma_{\mathrm{hh}}$} & \multicolumn{1}{c}{$5.68\pm0.68$} & \multicolumn{1}{c}{$-24.09\pm0.72$} \\ 
    & {[25,50]}
    & \multicolumn{1}{c}{$v_{\mathrm{hh}}$} & \multicolumn{1}{c}{$5.31\pm0.70$} & \multicolumn{1}{c}{$-26.07\pm0.75$} \\ 
    & $k=2$ & \multicolumn{1}{c}{$\sigma_{\mathrm{hh}}$} & \multicolumn{1}{c}{$5.47\pm0.68$} & \multicolumn{1}{c}{$-23.72\pm0.72$} \\ 
    \hline
  \end{tabular}
\caption{Linear regression fitting results of $\Delta R^v_{\mathrm{h}k}$ and $\Delta R^\sigma_{\mathrm{h}k}$ corresponding to different halo virial mass ranges $M_{\mathrm{vir}}$.}
\label{tab:fitting-results-hh}
\end{table*}

For light halos, in the intermediate region $I_{1}$, we observe that, by varying $\beta$ ($<0$) and $\lambda$ in the same direction, they exert opposing effects on $v_{\mathrm{hh}}$, as seen in the best-fit coefficients of the former $C^{v}_{\beta\mathrm{h}1} (\%) = 6.00 \pm 0.73$ and the latter $C^{v}_{\lambda\mathrm{h}1} (\%) = -25.79 \pm 0.77$. This trend persists in the two-halo regime ($I_2$), but the influence of $\lambda$ increases, with $C^{v}_{\lambda\mathrm{h}2} (\%) = -27.55\pm0.75 $, while the other coefficients remain largely unchanged. This contrasts with $\sigma_{\mathrm{hh}}$, where both effects are largely scale independent.

When sampled over heavy halos, $v_{\mathrm{hh}}$ and $\sigma_{\mathrm{hh}}$ exhibit similar sensitivity to both parameters as their light halo counterparts across both regions. Furthermore, the effect of $\lambda$ strengthens at larger scales for  $v_{\mathrm{hh}}$. This is evident in the transition from the $I_1$ to the $I_2$ region, where the coefficient for $v_{\mathrm{hh}}$ changes from $C^{v}_{\lambda\mathrm{h}1} (\%) = -24.61 \pm 0.76$ to $C^{v}_{\lambda\mathrm{h}2} (\%) = -26.07 \pm 0.75$.



Lastly, in a cross comparison between $v_{\mathrm{hh}}$ and $\sigma_{\mathrm{hh}}$, the former is more sensitive to variations in $\lambda$ than the latter at the two-halo regime for both halo mass ranges studied.


In summary, the mean halo-halo peculiar pairwise velocity $v_{\mathrm{hh}}$ in the two-halo regime provides a robust probe for constraining $\beta$ and $\lambda$ for heavy halos. Similarly, for light halos, $v_{\mathrm{hh}}$ remains sensitive to both parameters across the intermediate and two-halo regions. These results collectively demonstrate the discriminative power of the velocity statistics for studying interacting scalar field DM across various mass ranges and scale regimes.

\section{Conclusions} \label{sec:conclusions}

In this paper, we conduct a quantitative study to investigate how the coupling constant $\beta$ ($ < 0 $) and the scalar field potential slope $\lambda$ ($ > 0.6 $) in the Type 3 model influence the mean particle-particle (halo-halo) peculiar pairwise velocity $v_{\mathrm{pp}}$ ($v_{\mathrm{hh}}$) and mean peculiar velocity dispersion $\sigma_{\mathrm{pp}}$ ($\sigma_{\mathrm{hh}}$), which are sensitive proxies for structure formation \cite{zhang2024}. The observed effects stem from three main factors: the refitting of the cosmological parameters, alterations to the cosmological friction force and gravity acting on DM, and modifications to the cosmological expansion history.

Our findings reveal that the degeneracy between $\beta$ ($<0$) and $\lambda$ can be resolved using pairwise velocity statistics. Our key conclusions are summarized below:

\hspace*{\fill}

(i) For both the particle-particle and halo-halo pairwise velocity statistics, increasing $\beta$ (decreasing $|\beta|$, weaker coupling) and $\lambda$ (steeper potential) have opposite effects on the mean peculiar pairwise velocity. At $r>3\,\mathrm{Mpc}$, a weaker coupling increases the mean peculiar pairwise velocity magnitudes, while a steeper potential decreases them.

(ii) For $r<3\,\mathrm{Mpc}$, however, the same variation of $\beta$ ($\lambda$) would instead suppress (enhance) $v_{\mathrm{pp}}$. This reversal strongly suggests that shell crossing occurs at $r\sim3\,\mathrm{Mpc}$, marking the boundary between the nonlinear and linear regime. The opposite trend observed in the nonlinear regime arises from the orbital motions of DM particles, where friction enhances the shrinking of orbits. Since $|v_{\mathrm{pp}}|$ directly reflects the extent of structure formation, our results are consistent with those of \cite{palma2023}.

\hspace*{\fill}


In our current study, due to the limitations of the simulation resolutions, our results exhibit noticeable statistical noise at separation scales $r \leq 3\,\text{Mpc}$, especially for the halo-halo analysis. Therefore, future studies would benefit from higher-resolution simulations of the Type 3 model to better probe the velocity statistics at smaller scales.

Lastly, void properties, such as number density \cite{Pisani2015} and morphology \cite{Bos2012, Biswas2010}, could provide additional constraints on dark sector interactions, given their sensitivity to the dynamical DE EOS  \cite{Pisani2015, Bos2012, Biswas2010}.

\section*{Acknowledgements} \label{sec:acknowledgements}
We would like to thank Jiajun Zhang for the helpful advices and feedbacks on performing the N-body simulations and Thejs Brinckmann for the helpful guidance on utilizing \texttt{MontePython} \cite{Audren2013, Brinckmann2019}. This research is partially funded through grants awarded by the Research Grants Council of the Hong Kong Special Administrative Region, China (Project No.~14301214 and No.~AoE/P-404/18). We also extend our gratitude to the CUHK Central High Performance Computing Cluster for providing the computational resources used in this study.

\section*{DATA AVAILABILITY}  \label{sec:DATA AVAILABILITY}
The data that support the findings of this article are not publicly available upon publication because it is not technically feasible and/or the cost of preparing, depositing, and hosting the data would be prohibitive within the terms of this research project. The data are available from the authors upon reasonable request.


\bibliographystyle{apsrev4-2}
\nocite{*}


%


\appendix
\renewcommand{\appendixname}{APPENDIX}
\renewcommand{\thesection}{\Alph{section}}
\label{sec:appendix}
\section{\MakeUppercase{Supplementary material: Lagrangian and field equations}}
\label{sec: equations}

This section expands on the Lagrangian formulation of the Type 3 model introduced in Sec.~\ref{sec:develop}, providing derivations of the field equations and their connection to the continuity equations, which govern the scalar field and CDM fluid dynamics.


From the energy-momentum tensor components in Eq.~\eqref{Energy_Tensor}, the background energy density and pressure of the scalar field are derived as \cite{pourtsidou2013},
\begin{equation}
\rho_\phi = \left( \frac{1}{2} - \beta \right) \dot{\phi}^2 + V(\phi), 
\end{equation}
\begin{equation}
P_\phi = \left( \frac{1}{2} - \beta \right) \dot{\phi}^2 - V(\phi). 
\end{equation}

\begin{figure}
    \includegraphics[width=.95\columnwidth]{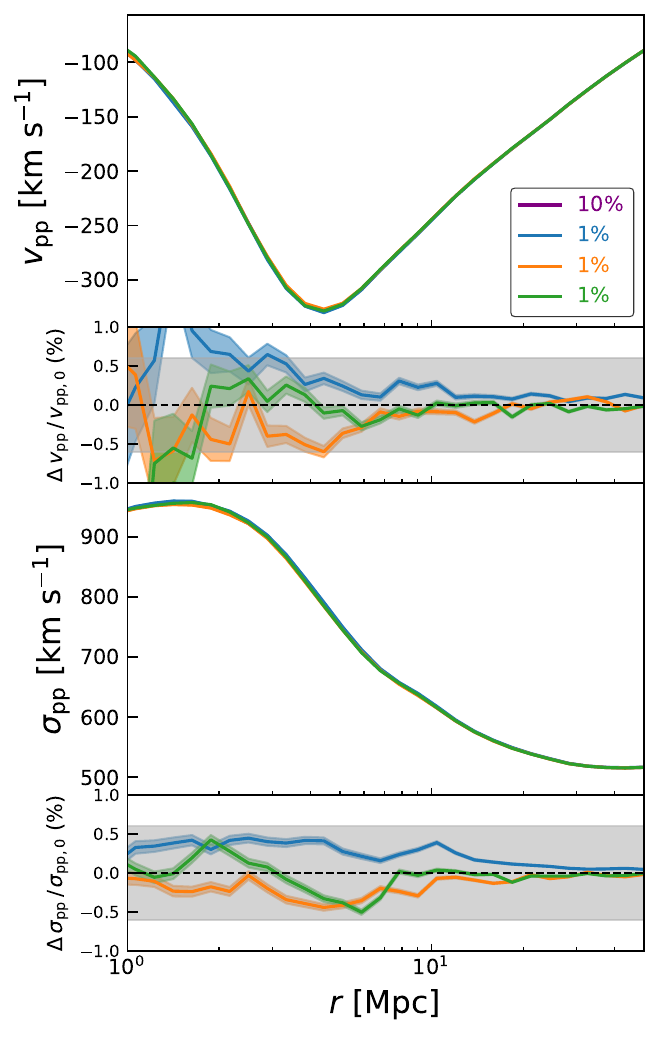}
    \caption{Mean particle-particle peculiar pairwise velocity $v_{\mathrm{pp}}$ (upper panel) and mean peculiar particle-particle pairwise velocity dispersion $\sigma_{\mathrm{pp}}$ (lower panel). The colors in the plots correspond to measurements using different fractions of randomly selected CDM particles from the simulations. The subpanels below the mean peculiar pairwise velocity and velocity dispersion plots show their fractional deviations relative to measurements in which $10\%$ of CDM particles have been selected. }
    \label{fig:CDM-fraction}
\end{figure}

\begin{figure*}
    \includegraphics[width=.85\textwidth]{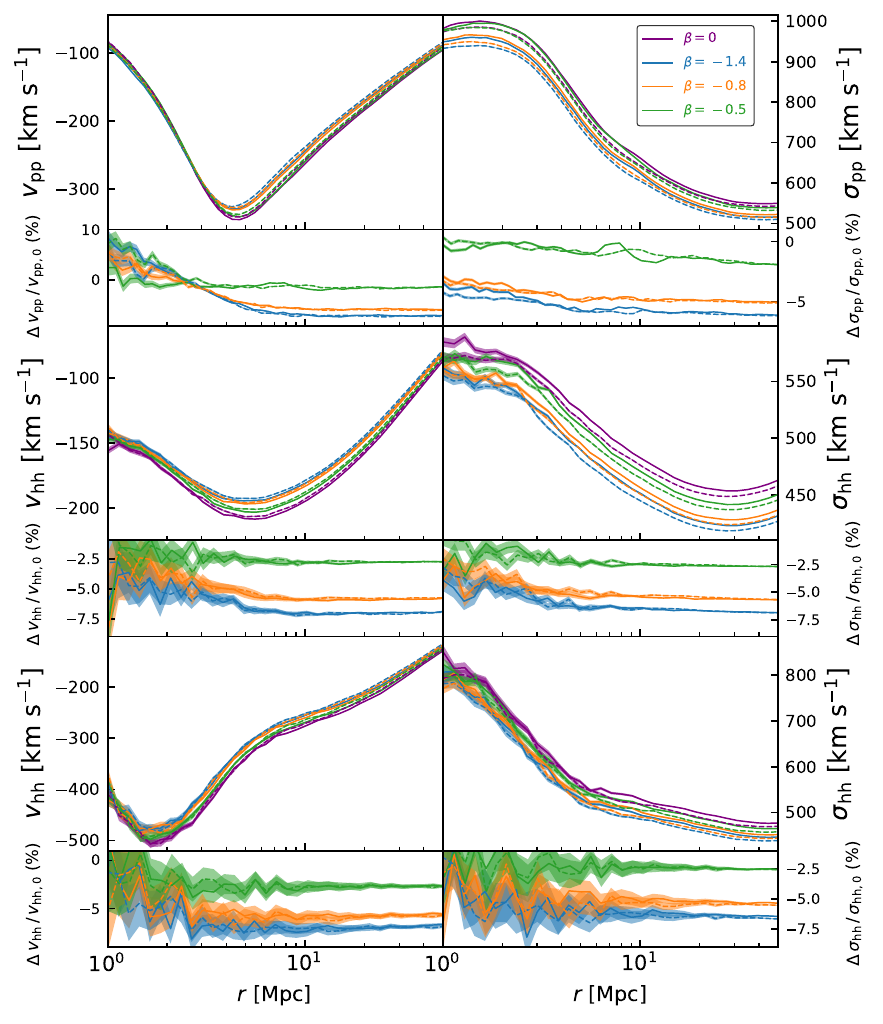}
    \caption{Mean particle-particle peculiar pairwise velocity $v_{\mathrm{pp}}$ (upper left panel) and velocity dispersion $\sigma_{\mathrm{pp}}$ (upper right panel). The middle and lower panels depict the mean halo-halo peculiar pairwise velocity $v_{\mathrm{hh}}$ (left) and mean peculiar velocity dispersion $\sigma_{\mathrm{hh}}$ (right) for light halos ($[10^{11}, 10^{13}]\,M_{\odot}$) and heavy halos ($[10^{13}, 10^{15}]\,M_{\odot}$), respectively. Solid and dashed lines represent measurements from random seeds 1 and 16180, respectively. The mean peculiar pairwise velocity and velocity dispersion for the uncoupled case (A0) are shown (purple lines) alongside those with different $\beta$ values at a fixed $\lambda = 0.6$. Subpanels below display the fractional deviations of the mean peculiar pairwise velocity and velocity dispersion relative to those of the fiducial (A0).}
    \label{fig:Cosmic Variance}
\end{figure*}

Varying Eq.~\eqref{eq:lagrangian} with respect to $\phi$ then leads to the continuity equation for the scalar field \cite{pourtsidou2013}
\begin{equation}
\nabla_\mu (F_Y \phi^\mu + F_Z u^\mu) - F_\phi = 0.  \label{scf_eq}
\end{equation}

This equation can be combined with Eq.~\eqref{eq:lagrangian} to derive Eq.~\eqref{eq: scf expli equation}, its explicit form.

Finally, using  Eq.~(\ref{Energy_Tensor}), the coupling current $J_{\mu} = -\nabla_{\nu}T^{\nu}_{(\phi)\mu} $ reads 
\begin{equation}
J_\mu = q^{\gamma}_{\mu} \bigl( \nabla_\nu ( F_{Z} u^\nu ) \nabla_{\gamma} \phi  + F_{Z} \nabla_\gamma Z + Z F_{Z}u^{\nu} \nabla_{\nu} u_{\gamma} \bigl) \label{coupling}
.\end{equation}

Here, $q^{\gamma}_{\mu} = u^{\gamma}u_{\mu} + {\delta^{\gamma}}_{\mu}$ is the metric projection operator. Notably, the terms in Eq.~\eqref{coupling} involve derivatives of at least first order (e.g.,$\nabla_\nu ( F_{Z} u^\nu)$, $\nabla_{\gamma} \phi$), meaning they vanish in the homogeneous background (zeroth order). Consequently, $J_{\mu} = 0$ at the background level. However, as shown in Eq.~\eqref{eq:euler eq}, pure momentum exchange is evident at the linear perturbation level through the modified Euler equation.

\section{\MakeUppercase{Supplementary material: Pairwise velocity dispersion}}
\label{sec: BBGYK equations}
This Section details the derivation of the approximations for the mean peculiar pairwise velocity dispersion $\sigma_{12}(r)$. 

Given a two-body interacting system (e.g. a particle pair), its dynamics can be captured by the second BBGKY equation, which describes the two-particle distribution function \cite{Peebles2020}. It is then integrated over the momentum space and simplified to give 
\begin{eqnarray}
\frac{\partial}{\partial t} \left(1 + \xi(r)\right) v_{12}^i  + \frac{\dot{a}}{a} \left(1 + \xi(r)\right) v_{12}^i  \nonumber \\+
 \frac{1}{a}  \sum_{j}^{} \frac{\partial}{\partial r^j} \left(1 + \xi(r)\right) \langle v_{12}^{i} v_{12}^{j} \rangle \nonumber \\ 
+ \frac{2G m}{a^2} \frac{r^i}{r^3} (1 + \xi(r)) + 2Gm \bar{n}  a \frac{r^i}{r^3} \int_{0}^{r} d^3 \mathbf{r} \xi(r) \nonumber \\ 
+ 2G m\bar{n} a \int_{0}^{r} d^3 \mathbf{r}_{3} \zeta(1, 2, 3) \, \frac{r_{13}^i}{r_{13}^{3}} = 0. \label{BBGKY}
\end{eqnarray}

Here, $m$ is the particle mass, $ \textbf{r}_{3}$ is the position vector of the third particle, $\textbf{r}_{13}=\textbf{r}_1-\textbf{r}_3$ and $\zeta(1, 2, 3)$ is the three-point correlation function. Just as the mean peculiar pairwise velocity reflects the evolution of $\xi(r,a)$ in Eq.~\eqref{vel_two-point}, Eq.~\eqref{BBGKY} shows how the mean peculiar velocity dispersion and gravity are related to the evolution of  $\zeta(1, 2, 3)$. The fourth term in the equation describes the attraction between the particle pair, while the last two terms account for the gravitational interaction with all the neighboring particles. For strong clustering, the last two terms become the dominant gravitational sources. At large separations, it is primarily the second to last term that provides gravity  ($\zeta \ll \xi \ll1 $). 

In Eq.~\eqref{BBGKY}, the term $\langle v_{12}^{i} v_{12}^{j} \rangle$ has a close relation to the mean velocity dispersion $\sigma_{12}$, and is shown to be 
\begin{equation}
\label{eq: approximation}
 \langle v_{12}^{i} v_{12}^{j} \rangle = \left(\frac{2}{3}\langle v^2 \rangle + \Sigma\right) \delta^{ij} + (\Pi - \Sigma) r^{i} r^{j} / r^2,
\end{equation}
where $\frac{2}{3}\langle v^2  \rangle$ represents the mean squared peculiar velocity of particles without correlated motions. Here, the factor of 2 comes from the pairwise coupling, while the factor of $\frac{1}{3}$ reflects the assumption of an isotropic Universe, and that the peculiar velocity variance is equally distributed along the three spatial dimensions. On the other hand, $\Pi$ and $\Sigma$ describe the influence of correlated motions on the velocity dispersion in the parallel and perpendicular components to the direction of interparticle separations. Hence, the projection of the velocity dispersion along the separation is simply given by 
\begin{equation}
\label{eq: sigma12 los}
\sigma^{2}_{12}(r) = \left(\frac{2}{3}\langle v^2 \rangle + \Pi\right).
\end{equation}

Based on Eq.~\eqref{BBGKY} and the condition ($\zeta \gg \xi \gg 1$) \cite{Mo2010, Peebles2020, Mo1997}, the cosmic virial theorem (CVT) can be derived to approximate $\sigma_{12}^{2}(r)$ at different separation scales.

\section{ \MakeUppercase{MCMC constraints}} 
\label{sec: mcmc constraints}
Table~\ref{tab:cosmo_params} shows the cosmological parameters for the Type 3 model obtained in \cite{Pourtsidou2025}, with the prior ranges $\beta \in [-2,0.5]$ and $\lambda \in [0,2.1]$. The parameters values of $\beta$ and $\lambda$ within the lower and upper bounds of the constraints are used to perform the MCMC refitting for the different scenarios, as detailed in Table~\ref{tab:mcmc_cases}. 

Additionally, prior to the release of the DESI DR2 BAO data, the Type 3 model was constrained using different datasets as shown by \cite{pourtsidou2016, linton2022}, with \texttt{MontePython} \cite{Audren2013, Brinckmann2019}. To ensure consistency, we use the same tool and datasets as \cite{pourtsidou2016, linton2022} to compare the performance of the Type 3 model and $\Lambda$CDM through the Bayes factor ($B$):

\begin{equation}
   B = \frac{K(\mathrm{T}3)}{K(\Lambda \mathrm{CDM})}. 
\end{equation}

Here, $K$ denotes the Bayesian evidence, computed using \texttt{MCEvidence} \cite{heavens2017}. We obtain $\log_{10} B = 1.04$, indicating that the Type 3 model is moderately favored over $\Lambda$CDM by the cosmological data considered. This finding provides another motivation for our study of the Type 3 model. Table~\ref{tab:cosmo_params_original} summarizes the cosmological parameters for both $\Lambda$CDM and the Type 3 model, derived under the prior ranges $\beta \in [-0.5, 0]$ and $\lambda \in [0, 2.1]$ \cite{pourtsidou2016, linton2022}. The table is organized into three sections: the upper panel lists the free parameters, the middle panel shows derived parameters computed from these, and the bottom panel reports the Bayesian evidence $\log_{10} K$.


\begin{table}
    \centering
    \begin{tabular}{lcc}
    \hline
    \hline
    Parameter & $\Lambda\text{CDM}$ & $\text{T3}$ \\
    \hline
    $n_s$ & $0.9672 \pm 0.0034$ & $0.969 \pm 0.004$ \\
    $100\theta_s$ & $1.04190 \pm 0.00023$ & $104.20 \pm 0.02$ \\
    $\Omega_b h^2$ & $0.02229 \pm 0.00012$ & $0.0223 \pm 0.0001$ \\
    $\Omega_c h^2$ & $0.11800 \pm 0.00061$ & $0.117 \pm 0.001$ \\
    $\tau_{\mathrm{reio}}$ & $0.0582^{+0.0064}_{-0.0073}$ & $0.06 \pm 0.01$ \\
    $H_0$ & $68.02 \pm 0.28$ & $66.9 \pm 0.6$ \\
    $\sigma_8$ & $0.8059 \pm 0.0057$ & $0.777^{+0.026}_{-0.015}$ \\
    $\lambda$ & $...$  & $1.00 \pm 0.4$ \\
    $\beta$ & $...$  & $-0.8^{+1.0}_{-0.6}$ \\
    \hline
    \end{tabular}
    \caption{Best-fit values (68\% CL) of the cosmological parameters for the $\Lambda$CDM and Type 3 model presented in \cite{Pourtsidou2025}, obtained by fitting with the Planck 2018 CMB, lensing \cite{planck2020b, Rosenberg2022, Carron2022, Aghanim2019}, DESI DR2 BAO \cite{AbdulKarim2025}, and the DES-Y5 Type Ia supernova sample\cite{DESCollaboration2024}. Parameter values of $\beta$ and $\lambda$ within the lower and upper bounds are used to perform the MCMC refitting, as shown in Table~\ref{tab:mcmc_cases}.}
    \label{tab:cosmo_params}
\end{table}

\begin{table}
    \centering
    \begin{tabular}{ccc}
    \hline
    \hline
    Parameter & $\Lambda\text{CDM}$ & $\text{T3}$ \\
    \hline     
    $\Omega_bh^2$ & $0.02262^{+0.00012}_{-0.00012}$ & $0.02256^{+0.00012}_{-0.00011}$ \\
    $\Omega_ch^2$ & $0.11700^{+0.00070}_{-0.00073}$ & $0.11747^{+0.00073}_{-0.00066}$ \\
    $100\theta_\mathrm{MC}$ & $1.0420^{+0.0003}_{-0.0003}$ & $1.04205^{+0.00025}_{-0.00024}$ \\
    $\tau_{reio}$ & $0.0492^{+0.0057}_{-0.0056}$ & $0.0554 \pm{0.0053}$ \\
    $\ln(10^{10}A_s)$ & $3.0301^{+0.0107}_{-0.0105}$  & $3.0392^{+0.0097}_{-0.0100}$ \\
    $n_s$ &  $0.9719^{+0.0034}_{-0.0031}$ & $0.9705 \pm{0.0031}$ \\
    $\beta$ & $...$  & $-0.453^{+0.093}_{-0.047}$ \\
    $\lambda$ &  $...$  & $1.21^{+0.167}_{-0.157}$ \\
    \hline
    $H_0$ & $68.74^{+0.33}_{-0.29}$ & $66.37^{+0.63}_{-0.62}$ \\
    $\Omega_m$ & $0.2969^{+0.0038}_{-0.0042}$ & $0.3180 \pm{0.0067}$ \\
    $\sigma_8$ & $0.7963\pm{0.0045}$ & $0.7626\pm{0.0093}$ \\
    $z_{reio}$ & $7.05^{-0.61}_{-0.55}$ & $6.64\pm{0.62}$ \\
    $10^9A_s$ & $2.0701^{-0.0223}_{-0.0215}$ & $2.0889\pm{0.2063}$ \\
    \hline
     $\mathrm{log_{10}}\,K$ & $-775.11$ & $-774.07$ \\
    \hline
    \end{tabular}
    \caption{Mean values (68\% CL) of the cosmological parameters for $\Lambda$CDM and the Type 3 model obtained by fitting with the Planck 2018 CMB, lensing \cite{Aghanim2020,Aghanim2019}, Planck SZ \cite{Aghanim2020, ade2014, Ade2015}, BOSS BAO \cite{Anderson2012}, and JLA \cite{Betoule2014} data. }
    \label{tab:cosmo_params_original}
\end{table}

\section{\MakeUppercase{Linear matter power spectrum}} 
\label{sec: Power spectrum}

\begin{figure}
    \includegraphics[width=.95\columnwidth]{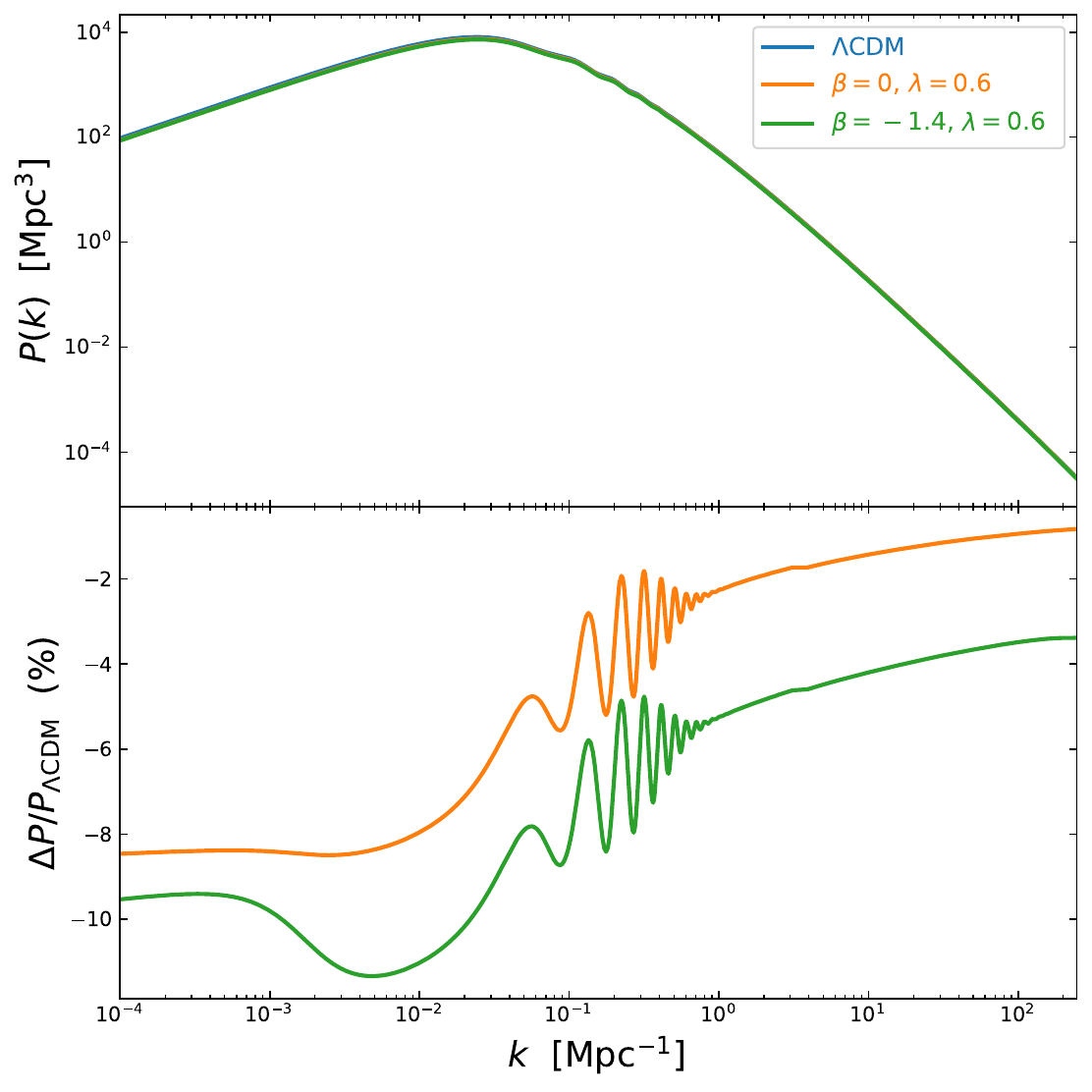}
    \caption{Linear matter power spectrum for $\Lambda$CDM (blue) and the Type 3 model with zero interaction (orange) and nonzero interaction (green) at $\lambda = 0.6$. The bottom panel shows the fractional deviation of each spectrum relative to $\Lambda$CDM.}
    \label{fig:power spectrum}
\end{figure}

Fig. \ref{fig:power spectrum} shows the linear matter power spectrum for the $\Lambda$CDM model and the Type 3 model with zero and nonzero $\beta$, which correspond to the cases A0 and A1 with the cosmological parameters displayed in Table \ref{tab:cosmo_params}, respectively. Since $\Lambda$CDM is the scenario with $\beta = 0$ and $\lambda = 0$, the impact of both parameters can be assessed by comparing its spectrum with those of the Type 3 model without interaction (A0: $\beta = 0$, $\lambda = 0.6$) and with interaction (A1: $\beta = -1.4$, $\lambda = 0.6$). 

The figure clearly indicates that increasing the potential slope ($\lambda$) weakens the clustering of dark matter particles relative to $\Lambda$CDM, thereby suppressing the power spectrum. This trend is consistent with the effect of $\lambda$ on pairwise velocity statistics at linear scales. Similarly, introducing a net interaction, where dark matter particles lose momentum to the scalar field, further suppresses the power spectrum compared to $\Lambda$CDM, agreeing with the directional influence of $\beta$ observed in pairwise velocity statistics at the linear regime. 

Because the linear matter power spectrum effectively captures these impacts on structure formation, N-body simulations using \texttt{ME-Gadget-2} can be subsequently performed for nonlinear analysis.

\section{\MakeUppercase{Fraction of CDM particles}} 
\label{sec: CDM fraction}

Due to the large number of CDM particles ($1024^{3}$) present in our simulations, it would be computationally costly to calculate the pairwise velocity statistics by accounting for all the particles. We therefore investigate how these measurements depend on the fraction of randomly sampled CDM particles.

Fig.~\ref{fig:CDM-fraction} displays the pairwise velocity statistics $v_{\mathrm{pp}}$ and $\sigma_{\mathrm{pp}}$ for scenario A1 ($\beta = -1.4$, $\lambda = 0.6$) under four sampling conditions: three configurations use $1\%$ of the total CDM particles, while the last uses $10\%$.

Our analysis reveals that reduced particle sampling yields statistically consistent results. The maximum discrepancy between $1\%$ and $10\%$ sampling fractions remains below 0.6\% at all scales relevant to this study ($r > 4\,\text{Mpc}$). Based on this finding, we adopt 1\% particle sampling for all mean peculiar pairwise velocity calculations presented in this work.

\hfill

\section{\MakeUppercase{Cosmic variance}} 
\label{sec: Cosmic variance}
In our analysis, we used random seed 1 in the IC generator for Type 3 model simulations. To verify the robustness of scalar field parameter effects ($\beta$ and $\lambda$), we conducted additional simulations with a distinct random seed (16180). 

As shown in Appendix~\ref{sec: CDM fraction}, mean peculiar pairwise velocities were calculated using $1\%$ of randomly selected CDM particles. Fig.~\ref{fig:Cosmic Variance} displays fractional deviations of $[v/\sigma]_{\mathrm{pp}}$ and $[v/\sigma]_{\mathrm{hh}}$ from the uncoupled scenario A0, where solid and dashed lines correspond to results from seeds 1 and 16180, respectively.  Comparing the deviations of each scenario across different seeds, although one finds that the magnitudes of $[v/\sigma]_{\mathrm{pp}}$ and $[v/\sigma]_{\mathrm{hh}}$ differ slightly, their mean deviations from A0 $ \left( \frac{\Delta [v/\sigma]_{\mathrm{pp}}}{[v/\sigma]_{\mathrm{pp,0}}} \,\, \text{and} \, \frac{\Delta [v/\sigma]_{\mathrm{hh}}}{[v/\sigma]_{\mathrm{hh,0}}} \right)$ remain consistent at the scales of our interest, $r>4 \,\,\text{Mpc}$. Due to this consistency, we select random seed 1 to perform our simulations. 






\end{document}